\documentclass[fleqn,usenatbib]{mnras}

\usepackage{newtxtext,newtxmath}
\usepackage[T1]{fontenc}

\DeclareRobustCommand{\VAN}[3]{#2}
\let\VANthebibliography\thebibliography
\def\thebibliography{\DeclareRobustCommand{\VAN}[3]{##3}\VANthebibliography}

\usepackage{graphicx}	% Including figure files
\usepackage{amsmath}	% Advanced maths commands

\title[Electron-density measurement for AGN outflows]{Electron densities from [S\,II] lines significantly overestimate the impact of ionised AGN outflows}

\author[Luke R. Holden et al.]{
Luke R. Holden$^{1}$\thanks{E-mail: l.holden@herts.ac.uk},
Daniel J. B. Smith$^{1}$,
Marina I. Arnaudova$^{2}$,
Clive. N. Tadhunter${^3}$,
\newauthor
\;Cristina Ramos Almeida$^{4, 5}$,
Shravya Shenoy$^{1}$,
Pedro H. Cezar$^{4, 5}$,
Soumyadeep Das${^1}$,
\newauthor
\;Akshara Binu${^1}$
\\
$^{1}$Centre for Astrophysics Research, University of Hertfordshire, Hatfield, AL10 9AB, United Kingdom. \\
$^{2}$Institute for Astronomy, University of Edinburgh, Royal Observatory, Blackford Hill, Edinburgh, EH9 3HJ, UK \\
$^{3}${School of Mathematics and Physical Sciences, University of Sheffield, Hounsfield Road, Sheffield, S3 7RH, United Kingdom} \\
$^{4}${Instituto de Astrofísica de Canarias, Calle Vía L\'actea, s/n, E-38205, La Laguna, Tenerife, Spain} \\
$^{5}${Departamento de Astrof\'isica, Universidad de La Laguna, E-38206, La Laguna, Tenerife, Spain}
}

\date{Accepted XXX. Received YYY; in original form ZZZ}

\pubyear{2025}

\begin{document}
\label{firstpage}
\pagerange{\pageref{firstpage}--\pageref{lastpage}}
\maketitle

\begin{abstract}
	To explain the properties of the local galaxy population, theoretical models require active galactic nuclei (AGN) to inject energy into host galaxies, thereby expelling outflows of gas that would otherwise form stars. Observational tests of this scenario rely on determining outflow masses, which requires measuring the electron density ($n_e$) of ionised gas. However, recent studies have argued that the most commonly used diagnostic may underestimate electron densities (and hence overestimate outflow masses) by several orders of magnitude, casting doubt as to whether ionised AGN-driven outflows can provide the impact needed to reconcile observations with theory. Here, we investigate this by applying two different electron-density diagnostics to Sloan Digital Sky Survey (SDSS) spectroscopy of the Quasar Feedback (QSOFEED) sample of 48 nearby type-2 quasars. Accounting for uncertainties, we find that outflow masses implied by the transauroral-line electron-density diagnostic are significantly lower than those produced by the commonly-used `strong-line' [S\,II](6717/6731) method, indicating a different origin of these emission lines and suggesting that these doubts are justified. Nevertheless, we show that it is possible to modify the [S\,II](6717/6731) electron-density diagnostic for our sample by applying a correction of $\mathrm{log}_{10}(n_{e\mathrm{,\, outflow}}\mathrm{ [cm}^{-3}\mathrm{]}) = \mathrm{log}_{10}(n_{e\mathrm{,[S\,II]}}\mathrm{ [cm}^{-3}\mathrm{]}) + 0.75(\pm0.07)$ to account for this, which results in values that are statistically consistent with those produced using the transauroral-line method. The techniques that we present here will be crucial for outflow studies in the upcoming era of large spectroscopic surveys, which will also be able to verify our results and broaden this method to larger samples of AGN of different types.
\end{abstract}

\begin{keywords}
galaxies: active -- galaxies: ISM -- galaxies: evolution -- galaxies: quasars: general -- ISM: clouds -- ISM: jets and outflows
\end{keywords}

%%%%%%%%%%%%%%%%%%%%%%%%%%%%%%%%%%%%%%%%%%%%%%%%%%

%%%%%%%%%%%%%%%%% BODY OF PAPER %%%%%%%%%%%%%%%%%%

\section{Introduction}
\label{section: introduction}

Electron density ($n_e$) is a key parameter in nebular astrophysics: it is crucial for calculating ionised gas masses (e.g. \citealt{MinkowskiAller1954, Rupke2005, Collins2009, Veilleux2020, Revalski2021, Arnaudova2024b}), temperatures (e.g. \citealt{Shaw1995, Osterbrock2006}), metallicities and cooling rates (e.g. \citealt{KewleyDopita2002}). As a result, it is also essential in our understanding of shock- and photoionisation (e.g. \citealt{Binette1996, Allen2008, Sutherland2017, Ferland2017, Meenakshi2022a}). 

In particular, electron densities play a major role in observational studies of gas outflows accelerated by active galactic nuclei (AGN; accreting supermassive black holes in the centres of galaxies), in which they are used to calculate the masses of warm-ionised gas ($10,000<T_e<25,000$\;K; e.g. \citealt{Nesvadba2006, Holt2011, Fiore2017, Harrison2014, Tadhunter2019}). These masses are then used to calculate parameters such as mass outflow rate and kinetic power, which are key metrics for quantifying the impact that outflows have on the star formation of their host galaxies, and hence their general importance in galaxy evolution. Commonly, such interpretations are made by comparing observationally-derived outflow properties to those used in models of galaxy evolution which invoke AGN-driven outflows (see discussions in \citealt{Harrison2018} and \citealt{Harrison2024}). Considering the large-scale observational studies that will be possible with upcoming and existing survey facilities, and given that recent simulation work is able to predict observable mass outflow rates and kinetic powers for different AGN types and outflow-acceleration mechanisms \citep{Mukherjee2018, Talbot2022, Ward2024}, being able to derive these properties from observations robustly is now especially important.

However, the true mass outflow rates and kinetic powers of AGN-driven outflows remain highly uncertain due to difficulties in measuring them observationally. Principally, this is because of limitations of the methods that are traditionally used to estimate electron densities. It has been argued that the most commonly-used density diagnostic, the [S\,II](6717/6731) flux ratio, may potentially underpredict true electron densities by several orders of magnitude \citep{Rose2018, Baron2019b, Revalski2022, Holden2023, HoldenTadhunter2023, Speranza2024}: the ratio is only strongly sensitive to densities in the range $2.0\lesssim\mathrm{log}_{10}(n_e \mathrm{[cm }^{-3}\mathrm{]})\lesssim3.5$, outside of which it becomes saturated. Thus, if the true electron densities of outflowing gas are above the upper limit of this range, the ratio will provide potentially significant underestimations.

Meanwhile, studies using alternative density diagnostics have demonstrated that AGN-driven ionised-gas outflows can have a wide range of densities ($2.0<\mathrm{log}_{10}(n_e \mathrm{[cm }^{-3}\mathrm{]})<7.0$: \citealt{Collins2009, Crenshaw2015, Baron2019b, Revalski2021, Revalski2022}). Since mass outflow rates and kinetic powers depend inversely on electron density (see \citealt{Harrison2018}), this indicates that ionised outflows may have a significantly reduced impact on host galaxies than would be expected based on electron-density measurements made with the [S\,II](6717/6731) ratio.

In particular, a technique first introduced by \citet{Holt2011} that makes use of the higher-critical-density transauroral \citep{Boyce1933} [O\,II]$\lambda\lambda7319,7330$ and [S\,II]$\lambda\lambda4068,4076$ emission lines has now been used to determine a similarly-wide range of electron densities for outflows in galaxies hosting AGN of various types, including Seyferts \citep{Davies2020, HoldenTadhunter2023}, quasars \citep{RamosAlmeida2019, Speranza2024, Bessiere2024}, ultraluminous infrared galaxies (ULIRGs; \citealt{Rose2018, Spence2018}), and compact radio galaxies \citep{Santoro2018, Santoro2020}. Moreover, using spatially-resolved observations of a nearby Seyfert\;2 galaxy, \citet{Holden2023} demonstrated that, for the same outflowing gas clouds, the densities measured by the [S\,II](6717/6731) ratio were half-an-order of magnitude lower than those measured with the transauroral-line method despite being within the sensitivity range of the former technique. In addition, \citet{RamosAlmeida2025} found that the densities derived from high-ionisation $\mathrm{[Ne\;V]}14.3, 24.3$\;\textmu m emission lines in five quasars were similar to those measured from the transauroral lines, but significantly larger than those produced by the [S\,II] ratio. Taken together, these results indicate a different physical origin for the transauroral and $\mathrm{[S\,II]}\lambda\lambda6717,6731$ lines.

Although the transauroral-line technique now finds regular use in observational studies of AGN-driven outflows, there has not yet been a study that focuses on comparing the densities derived from this method to those obtained using the commonly-used [S\,II](6717/6731) ratio for a large number of objects. To address this, here we perform statistical analyses of the electron-density values produced by both diagnostics for the 48 nearby type-2 quasars (QSO2s) of the \href{https://research.iac.es/galeria/cristina.ramos.almeida/qsofeed/}{QSOFEED} sample \citep{RamosAlmeida2022}. Using the transauroral lines and [S\,II](6717/6731) ratio, \citet{Bessiere2024} derived electron densities for this sample for use in mass-outflow-rate calculations, but did not perform a detailed comparison of the two techniques. Moreover, there are potentially significant uncertainties associated with modelling the faint transauroral emission lines, as well as the density-measurement methods themselves (which may incur errors of 0.2--0.7\;dex in the case of the transauroral-line technique: \citealt{Santoro2020}). Since this will impact direct comparisons between the two electron-density diagnostics, in this study, we use Markov Chain Monte Carlo (MCMC) emission-line fitting and develop a new Monte Carlo approach to electron-density measurement that allows us to ensure that our comparisons are sufficiently robust. In this way, we aim to accurately quantify the differences in the values derived from each technique, provide a firm basis for interpretations regarding the physical origin of the emission lines involved, and search for ways to improve electron-density measurements for future observational tests of the role of AGN-driven outflows in galaxy evolution.

This paper is structured as follows. In Section\;\ref{section: sample_and_observations}, we describe the sample of QSO2s and the spectroscopic observations; our method for fitting the spectra of these objects is detailed in Section\;\ref{section: spectral_fitting}. The approach we take to measuring electron densities using the two methods is given in Section\;\ref{section: electron_densities}, the results of which we present in Section\;\ref{section: results} and discuss in a broader context in Section\;\ref{section: discussion}. Finally, we give our conclusions in Section\;\ref{section: conclusions}.

\section{Sample and observations}
Our sample and observations are described in detail by \citet{Bessiere2024}, from which we provide a summary of the most relevant information here. 

\subsection{The QSOFEED sample}
\label{section: sample_and_observations}
\label{section: sample_and_observations: sample}

In this work we consider the complete QSOFEED \citep{RamosAlmeida2022, Bessiere2024} sample of 48 QSO2s. The QSOFEED project aims to quantify the properties of multiphase (gas of different temperatures and conditions; see \citealt{Cicone2018}) AGN-driven outflows and to determine their impact on their host galaxies (see \citealt{RamosAlmeida2019, RamosAlmeida2022, RamosAlmeida2023, RamosAlmeida2025, Speranza2022, Speranza2024, Audibert2023, Audibert2025, Bessiere2024, Holden2024, Zanchettin2025}).

The sample consists of all objects with $z<0.14$ and $L_\mathrm{[O\,III]}>10^{42}$\;erg\;s$^{-1}$ in the catalogue of Sloan Digital Sky Survey (SDSS: \citealt{York2000}) type-2 quasars presented by \citet{Reyes2008}, which itself was defined based on optical emission-line ratios, line widths, and [O\,III] equivalent width. The resulting objects have bolometric luminosities in the range $44.9<\mathrm{log}_{10}(L_\mathrm{bol} [\mathrm{erg\;s}^{-1}])<46$, high stellar masses ($10.6<\mathrm{log}_{10}(M_\star \mathrm{[M}{_\odot]})<11.7$), and $65^{+6}_{-7}$\;per\;cent are undergoing a merger event \citep{Pierce2023}. 

We chose this sample for our study because prior analysis of the warm-ionised-gas kinematics (derived from the [O\,III]$\lambda\lambda4959,5007$ doublet) by \citet{Bessiere2024} revealed that at least 85\;per\;cent of the objects present clear signatures of outflows in their spectra. Moreover, the spectra of the objects display prominent [S\,II]$\lambda\lambda4068,4076$, [S\,II]$\lambda\lambda6717,6731$, [O\,II]$\lambda\lambda3726,3729$ and [O\,II]$\lambda\lambda7319,7330$ emission lines.

\subsection{SDSS and BOSS spectra}
\label{section: sample_and_observations: observations}

Of the 48 targets that comprise the QSOFEED sample, spectra of 43 were taken from the SDSS legacy survey \citep{Abazajian2008}, with spectra for the remaining five targets taken from BOSS (Baryon Oscillation Spectroscopic Survey: \citealt{Dawson2013}) observations. The former spectra have a wavelength range of 3800--9200\;{\AA} and a spectral resolution of $R=1800$--2200, while the latter have a wavelength range of 3600--10000\;{\AA} and a spectral resolution of $R=1300$--2600. We note that, while the fibre diameters of the two spectrographs differ (3\;arcseconds for SDSS; 2\;arcseconds for BOSS), our analysis is not affected by this since we consider emission-line-flux ratios exclusively. However, we note that, due to emission-line contamination from star formation in the host galaxy, such aperture effects may be significant for future works that use samples of low-luminosity AGN.

The wavelength ranges for both datasets cover the emission lines required for the transauroral-line technique at the redshifts of the targets. However, the relatively-low spectral resolution prevents robust separation of broad (outflowing) and narrow (non-outflowing) components for all of the required emission-line profiles. Thus, we consider only total line fluxes in our analysis\footnote{Studies of higher spectral resolution have presented evidence that outflowing gas has higher electron densities than non-outflowing gas by 0.2--2.0\;dex (\citealt{Holt2011, Rose2018, Holden2023}; Holden et al. in prep; see also \citealt{Perna2017b} \& \citealt{Kakkad2018}). Since we do not separate outflowing and non-outflowing emission here, it is likely that our derived electron density values for both techniques are underestimates.}.

\section{Spectral fitting}
\label{section: spectral_fitting}

\subsection{Preparation of spectra}
\label{section: spectral_fitting: preparation}

To fit the spectra in our sample, we made use of the spectral-fitting code that we have developed for the upcoming WEAVE-LOFAR survey \citep{Smith2016} --- this code is based on that used by \citet{Arnaudova2024b}, Arnaudova et al. (submitted), and \citet{HoldenTadhunter2025}. First, the spectra were de-redshifted using the SDSS redshift values, and the extinction values from the dust maps of \citet{Schlegel1998} were used to correct each for Galactic extinction using the $R_v=3.1$ \citet{Fitzpatrick1999} extinction law; this was implemented using the \textsc{\href{https://extinction.readthedocs.io/en/latest/}{extinction} Python} module.

\subsection{Stellar continuum modelling and subtraction}
\label{section: spectral_fitting: stellar_continuum_modelling}

Since the transauroral [O\,II]$\lambda\lambda7319,7330$ and [S\,II]$\lambda\lambda4068,4076$ emission lines are typically faint relative to the continuum, we modelled and subtracted the stellar continuum for each object before performing the emission-line fits. We first masked regions corresponding to $\pm750$\;km\;s$^{-1}$ around all emission lines and then used the \textsc{pPXF} code \citep{Cappellari2004, Cappellari2017, Cappellari2023} with the \citet{Bruzual2003} stellar templates to fit the continuum in logarithmic-wavelength space. In many cases, a mixture of different templates was required to adequately fit the continuum. After the fits were completed, the resulting stellar-continuum model was interpolated onto the linear wavelength scale of the original spectrum. 

Key stellar features, such as the Ca\,II K, Mg\,I, and G-band absorption, were used to verify the accuracy of the fits (see example in Appendix\;\ref{appendix: ppxf}). After deeming the fits acceptable, we subtracted the modelled stellar continua from the spectra.

\subsection{Emission-line fitting}
\label{section: spectral_fitting: emission_line_fitting}

After subtracting the modelled stellar continua, we fit the lines required for our density diagnostics ([O\,II]$\lambda\lambda3726,3729$, [S\,II]$\lambda\lambda4068,4076$, [S\,II]$\lambda\lambda6717,6731$, [O\,II]$\lambda\lambda7319,7330$\footnote{Each line in the [O\,II]$\lambda\lambda7319,7330$ doublet is actually itself a doublet: [O\,II]$\lambda\lambda7319,7320$ and [O\,II]$\lambda\lambda7330,7331$. Considering that the wavelength separations of the lines in these doublets are far below the resolution of our spectra, we model them as single lines.}) simultaneously in velocity space using MCMC sampling. This is implemented in our emission-line fitting code using the \textsc{emcee} Python package \citep{FormanMackey2013}, which is an implementation of the Affine Invariant MCMC Ensemble sampler \citep{Goodman2010}. Since our models contain a large number of parameters in many cases, we used the Differential Evolution Markov Chain (DE-MC) algorithm described by \citet{terBraakVrugt2008} for efficient sampling of high-dimensionality posterior distributions.

The number of walkers in the MCMC chains was set to be twice the number of dimensions (i.e. parameters) of the model. At every 1000 steps, the autocorrelation time for each parameter was calculated from the latter half of its respective walker chain --- if all autocorrelation times were less than 5\;per\;cent of half of the current chain length then the chains were considered to be converged, and therefore the sampling was stopped.

The emission lines were modelled using $N_g$ Gaussian components, where $N_g$ is iteratively increased from one. For each iteration, the Bayesian Information Criterion (BIC) was used to quantify if the quality-of-fit had improved while accounting for the increased model complexity; following the recommendations of \citet{Raftery1995}, if the decrease in BIC was greater than 6 (i.e. $\mathrm{BIC}_{N_g} - \mathrm{BIC}_{N_{g+1}} > 6$) then the iteration was accepted, and the process was repeated until this criterion was not satisfied. In this way, we were able to describe the complex emission-line profiles that are expected for outflowing gas while avoiding overfitting; 45 objects (93.8\;per\;cent of the sample) required more than one Gaussian component in their emission-line models, 31 (64.6\;per\;cent of the sample) required three or more Gaussian components, and 5 (10.4\;per\;cent of the sample) required four Gaussian components.

\begin{figure*}
	\centering
	\includegraphics[width=\linewidth]{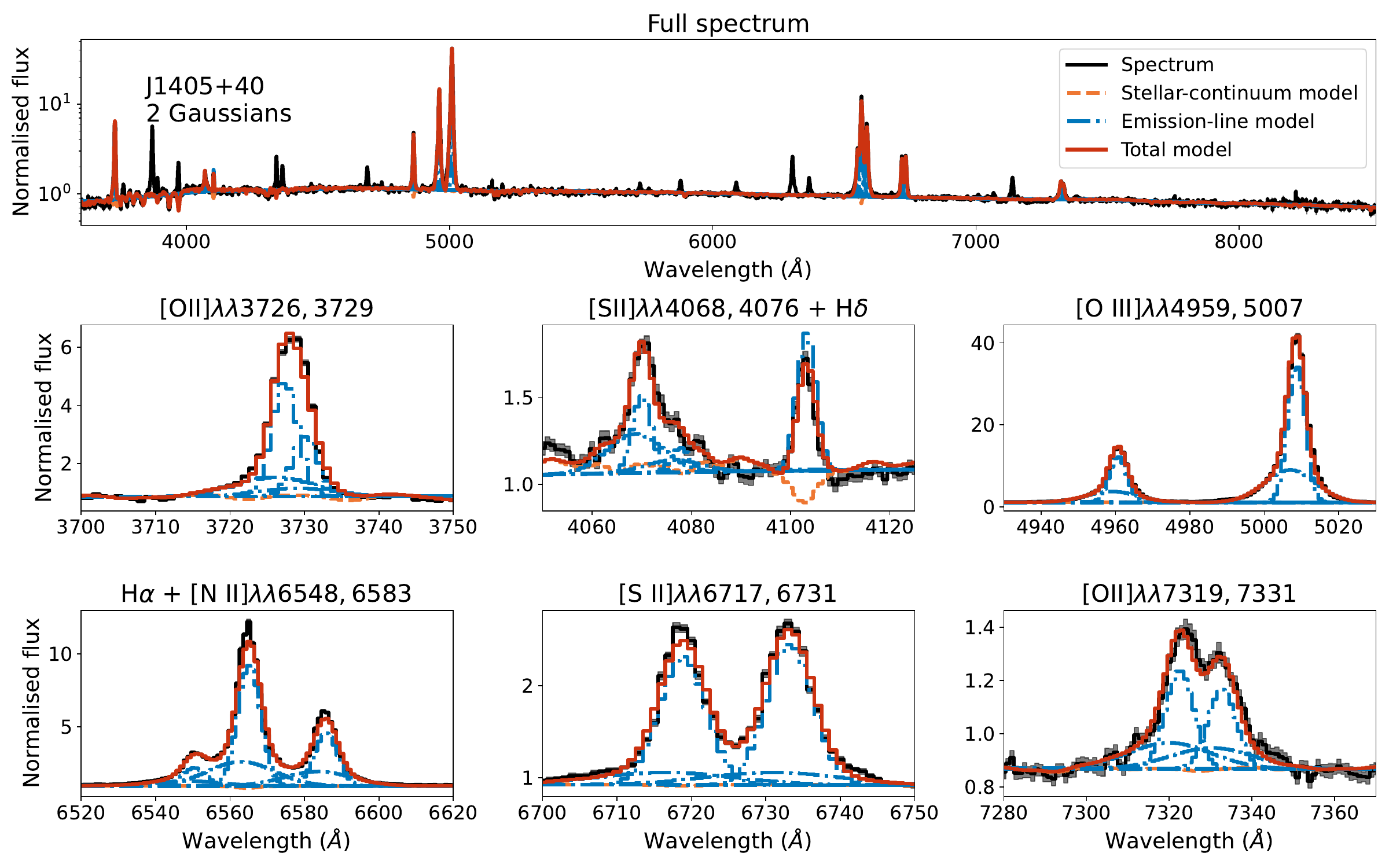}
	\caption{Spectral fit to the flux-normalised, de-redshifted spectrum of the QSO2 J1405+40 that was produced using our fitting routine. The observed spectrum is represented by the solid black line, with $1\sigma$ flux uncertainties shown in shaded grey; the overall fit to the spectrum is shown as a solid red line; the emission-line model is shown as a dashed-dotted blue line, and the stellar-continuum fit (see Section\;\ref{section: spectral_fitting: stellar_continuum_modelling} and Figure\;\ref{fig: stellar_continuum_fit}) is shown as a dashed orange line. The full wavelength range of the spectrum is shown in the top panel (in which the flux axis is shown logarithmically for presentation purposes); the other panels show the fits to key emission lines used in our analysis along with the $\mathrm{[O\,II]}\lambda\lambda4959,5007$ doublet and H$\alpha+\mathrm{[N\,II]}\lambda\lambda6548,6583$ lines that we include to better constrain the fits. Here, the emission-line models have been offset to the level of the continuum for presentation purposes.}
	\label{fig: emission_line_fit}
\end{figure*}

Priors for model parameters in the MCMC routine were physically motivated: for example, the Gaussian peak fluxes and widths were constrained to be positive, and the peak fluxes for Gaussian components within doublets were forced to be within the ratio ranges established by atomic physics ($0.41<\mathrm{[O\;II](3729/3726)}<1.50$; $3.01<\mathrm{[S\;II](4068/4076)}<3.28$, $0.46<\mathrm{[S\;II](6717/6731)}<1.45$; determined using the \textsc{PyNeb Python} package: \citealt{Luridiana2015}). While the velocity shift and width of a given Gaussian component was the same for all emission lines, the peak fluxes were allowed to vary within these priors. We note that we also included the bright H$\beta$, $\mathrm{[O\,III]}\lambda4959,5007$, and H$\alpha$+$\mathrm{[N\,II]}\lambda\lambda6548,6583$ lines in our fits (with fixed flux ratios of $1:2.99$ and $1:2.92$ for the $\mathrm{[O\,III]}\lambda4959,5007$ and $\mathrm{[N\,II]}\lambda\lambda6548,6583$ doublets, respectively) in order to better constrain the profiles of the fainter transauroral lines, although we do not consider these in our analysis. In order to ensure that the walkers converged to the overall regions of maximum probability as quickly as possible, their initial positions in parameter space were determined using a least-squares fit of the model for a given iteration, with bounds set to be the same physically-motivated constraints that were used for the MCMC priors. An example of the resulting fit to one of the objects in our sample using our spectral-fitting routine is shown in Figure\;\ref{fig: emission_line_fit}.

After fitting the spectrum for each object in this way, the emission-line fit parameters (the peak fluxes, centroid wavelengths, and widths of the Gaussians) were determined by taking the 50th percentiles of the marginalised posterior distributions, with $1\sigma$ uncertainties estimated based on the 16th and 84th percentiles. The summed total flux and associated uncertainty of all Gaussian components for each emission line, which we present in Table\;\ref{tab: fluxes}, were then calculated using these values. All of the required diagnostic emission lines/doublets are detected to at least the $3\sigma$ level, except for the transauroral [O\,II]$\lambda\lambda7319,7330$ doublet in one object (J1533+35).

\section{A Monte Carlo approach to electron-density measurement}
\label{section: electron_densities}

In addition to the flux uncertainties from the emission-line fits, there are several other sources of uncertainty that are involved when calculating electron densities using both the [S\,II](6717/6731) ratio and the transauroral-line technique presented by \citet{Holt2011}, which we discuss in this section. To ensure that our calculated values accounted for this, here we derive electron densities by generating 10,000 Monte Carlo realisations for each object using both methods.

\subsection{Electron densities from the [S\,II](6717/6731) ratio}
\label{section: electron_densities: sii}

\begin{figure}
	\includegraphics[width=\linewidth]{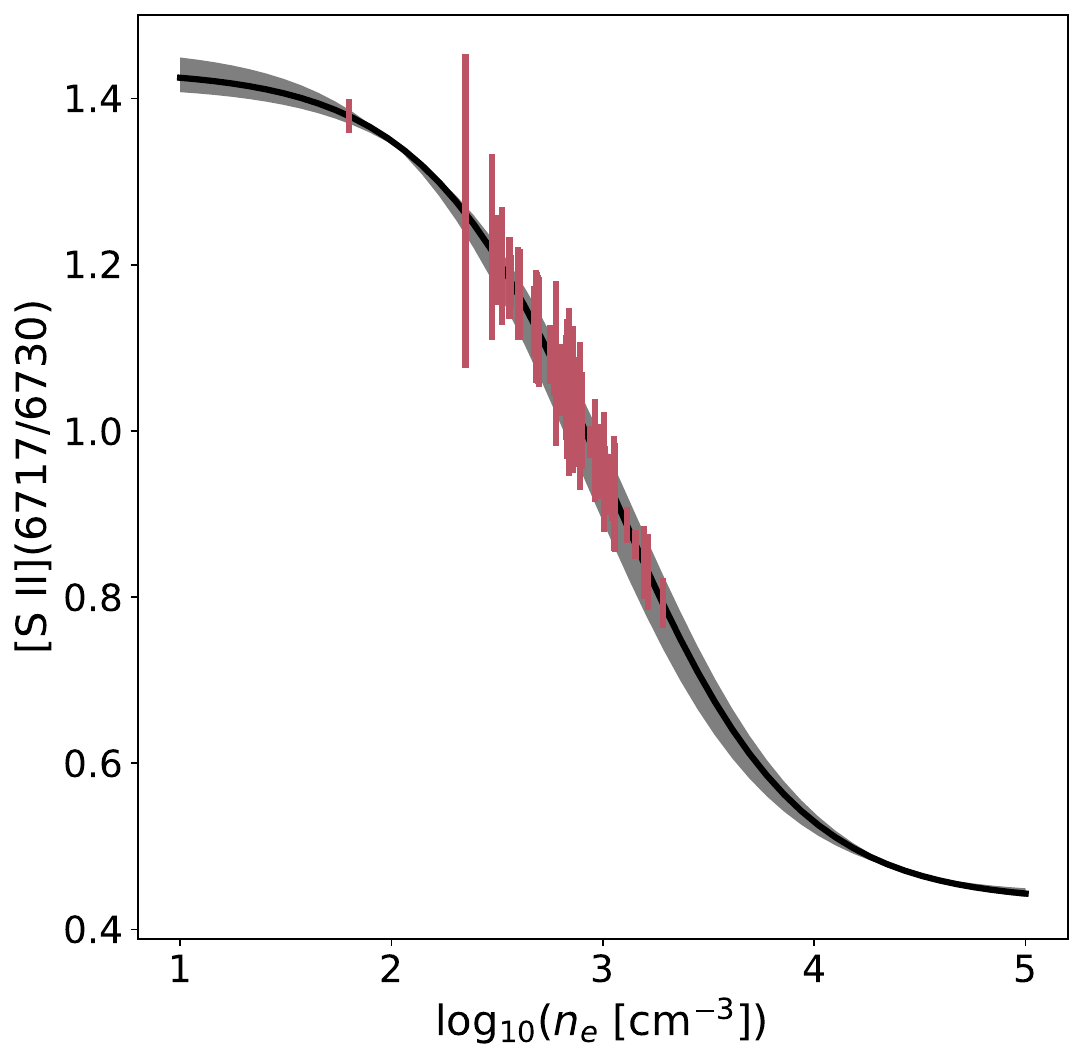}
	\caption{The variation of the [S\,II](6717/6731) flux ratio with electron density, as modelled using the \textsc{PyNeb} module for an electron temperature of $T_e=15\mathrm{,}000$\;K (black line); the shaded grey area shows the values of the ratio between $8000<T_e<22000$\;K. The solid red lines indicate the measured ratio with $1\sigma$ uncertainties for each object in the sample at its corresponding electron-density value.}
	\label{fig: sii_curve}
\end{figure}

The [S\,II](6717/6731) flux ratio is the most commonly used electron-density diagnostic in nebular and AGN-driven-outflow studies (see Chapter\;5 of \citealt{Osterbrock2006} for an introduction). It is strongly dependent on electron density between values of $2.0\lesssim\mathrm{log}_{10}(n_e \mathrm{[cm}^{-3}\mathrm{]})\lesssim3.5$, beyond which it becomes asymptotic (see Figure\;\ref{fig: sii_curve}). Therefore, by measuring the ratio of the fluxes of the two lines in the doublet and comparing this to the values expected from atomic physics, electron densities can be estimated. In Figure\;\ref{fig: sii_curve}, we show the [S\,II](6717/6731) flux ratio as a function of electron density along with the measured value for each of the objects in our sample, which we determined in the following way.

To calculate an electron density value for each Monte Carlo realisation, we began by randomly drawing [S\,II]$\lambda6717$ and [S\,II]$\lambda$6731 flux values from normal distributions with the means taken to be the values derived from the MCMC chains and the deviations taken to be the associated uncertainty. We then used these values to calculate a [S\,II](6717/6731) ratio for the realisation. To ensure that the resulting randomly-drawn ratio values were physical, we set any that fell outside the range $0.46<\mathrm{[S\,II]}(6717/6731)<1.45$ to its upper or lower limit, whichever was closest. 

Another source of uncertainty in calculating electron densities with this ratio is its weak dependence on electron temperature; this effect is shown as a shaded region in Figure\;\ref{fig: sii_curve}. Therefore, for each realisation, we randomly selected an electron-temperature value from a normal distribution of mean 15,000\;K and standard deviation 2000\;K, chosen to cover typical temperatures in the narrow-line regions (NLRs) of AGN \citep{Osterbrock2006}; these temperatures are comparable to those derived for six objects in the QSOFEED sample ($12,000<T_e<17,000$\;K: Cezar et al., in prep.). The randomly-chosen [S\,II](6717/6731) ratio and electron-temperature values were then used to determine an electron density for each realisation using the \textsc{PyNeb} python package; we combined all realisations to produce an electron-density distribution for each object.

\subsection{Electron densities from the transauroral-line technique}
\label{section: electron_densities: tr}

In addition, we measure electron densities for each object using the technique introduced by \citet{Holt2011}, which involves measuring two transauroral-line flux ratios ($TR$) and comparing them to the predictions of photoionisation modelling:
\begin{align}
	& TR(\mathrm{[O\,II]}) = F(3726 + 3729)/F(7319 + 7331), \label{eq: tr_oii}\\
	& TR(\mathrm{[S\,II]}) = F(4068 + 4076)/F(6717 + 6731). \label{eq: tr_sii}
\end{align}

\begin{figure}
	\centering
	\includegraphics[width=1\linewidth]{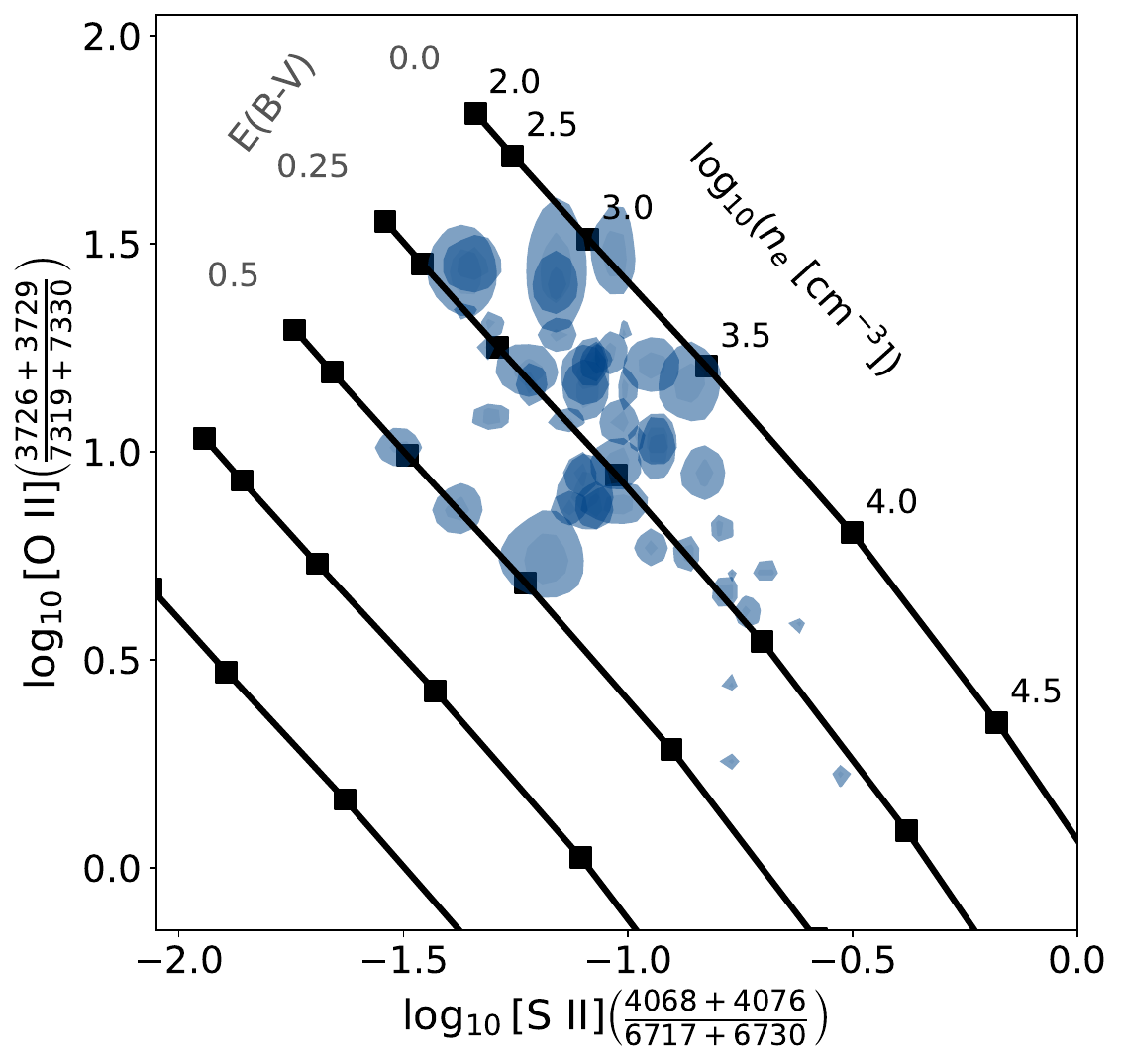}
	\caption{Example transauroral-line-ratio ($TR$) grid for an ionising-source spectral index of $\alpha=1.5$, an ionisation parameter of $\mathrm{log}U=-3.00$, and solar-metallicity gas; the black points represent the line ratios predicted by this photoionisation model for gas of different electron densities ($2.00<\mathrm{log}_{10}(n_e\mathrm{[cm}^{-3}])<6.00$) and colour-excess values ($0.01<E(B-V)<1.00$). The blue shaded regions contain 67\;per\;cent of the Monte Carlo realisations for each object in our sample.}
	\label{fig: tr_grid}
\end{figure}

We generated plane-parallel, radiation-bounded photoionisation models for gas with no dust depletion and of varying number density using version C23 of the \textsc{cloudy} code \citep{Chatzikos2023}. Since we have no direct prior information about the AGN spectral-energy-distributions (SEDs), ionisation parameters or gas metallicities of the objects in our sample, we generated a series of models in which we vary these parameters within reasonable ranges that are expected for the interstellar medium (ISM) under a variety of conditions\footnote{\citet{HoldenTadhunter2023} demonstrated that in the case of shock-ionised gas, an error of $\pm0.38$\;dex is induced in electron densities measured using $TR$ grids produced by photoionisation modelling. However, we highlight that this potential uncertainty is smaller than that induced by varying the photoionisation model parameters, and so we do not directly account for potential shock ionisation in this work.} --- this process is detailed in Appendix\;\ref{appendix: analytic_tr_ne}. For each model, we applied dust extinction to the simulated values of the transauroral-line-ratios using the $R_v=3.1$ extinction law from \citet{Cardelli1989} with colour excesses in the range $0.0<E(B-V)<1.0$. This process resulted in grids of simulated $TR$ values, an example of which is shown in Figure\;\ref{fig: tr_grid} as joined black squares.

Measured $TR$ values for our sample were calculated by randomly sampling from normal distributions based on the measured line fluxes and their associated uncertainties, as we did for the [S\,II]-derived densities in Section\;\ref{section: electron_densities: sii}. For each realisation, we randomly selected one of the photoionisation models that we generated and solved the analytical expression derived in Appendix\;\ref{appendix: analytic_tr_ne} to calculate electron-density values:

\begin{equation}
	\centering
	\begin{split}
		A\mathrm{log}_{10}n_e^3 + B\mathrm{log}_{10}n_e^2& + C\mathrm{log}_{10}n_e + D \\
				       & = \mathrm{log}_{10}TR(\mathrm{[O\,II]}) - m\mathrm{log}_{10}TR(\mathrm{[S\,II]})
	\end{split}
	\label{eq: tr_ne}
\end{equation}

\noindent
where $A$, $B$, $C$, $D$, and $m$ are constants, all of which (except $m$) depend on the parameters of the photoionisation model; the values of these constants for different model parameters are given in Table\;\ref{tab: tr_expression_constants}. 

For individual objects, we estimate electron-density values for each technique as the 50th percentile of the respective Monte Carlo density distributions, and the upper and lower uncertainties as the 16th and 84th percentiles --- we present these values in Table\;\ref{tab: ne}.

\begin{figure}
	\centering
	\includegraphics[width=\linewidth]{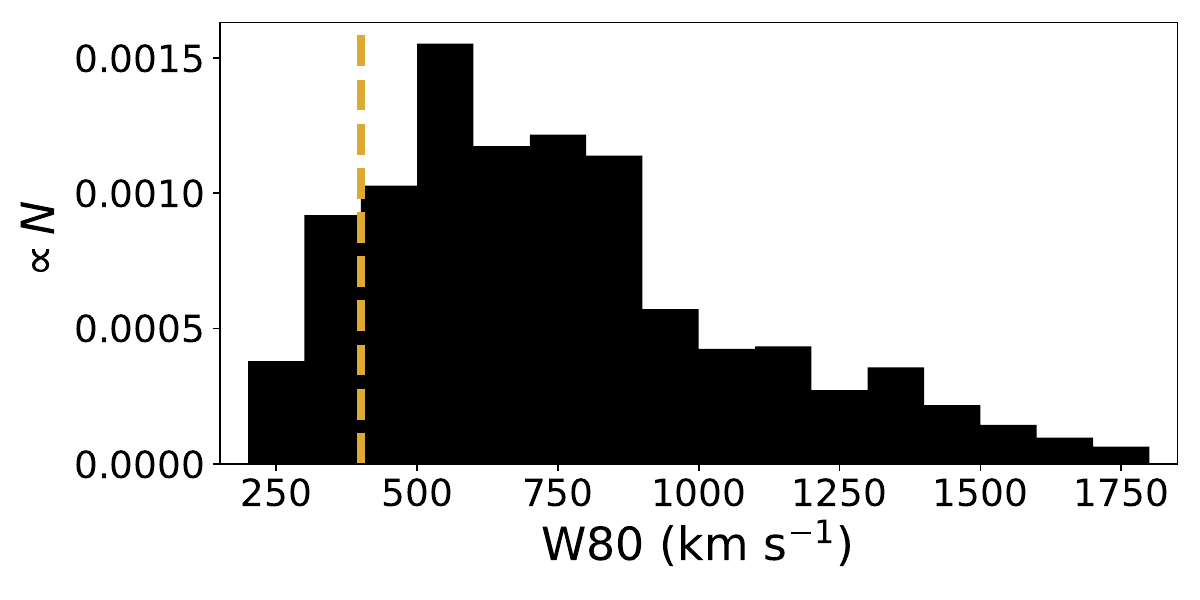}
	\caption{Non-parametric velocity width ($W_{80}$; the velocity width containing 80\;per\;cent of the total line flux) distribution for the $\mathrm{[O\,III]}\lambda5007$ emission-line profiles in our sample. The dashed yellow line corresponds to a line width of $400$\;km\;s$^{-1}$, above which the profiles can be considered to have a significant non-rotational component (see \citealt{Bessiere2024}).}
	\label{fig: w80_hist}
\end{figure}

To verify the extent to which the outflowing gas dominates the line profiles, for each Monte Carlo realisation, we also calculated the non-parametric velocity width for the line profile of $\mathrm{[O\,III]}\lambda5007$. This was done by measuring the velocities that contained 10 and 90\;percent of the total line flux ($v_{10}$ and $v_{90}$, respectively) and using these to calculate the velocity width that contains 80\;per\;cent of the line flux ($W_{80}=v_{90}-v_{10}$). We show the distribution of the velocity widths measured in this way for the entire sample in Figure\;\ref{fig: w80_hist}, from which it can be seen that high velocities dominate the line profiles; the mean value is $W_\mathrm{80}=797\pm60$\;km\;s$^{-1}$. This closely follows the findings of \citet{Bessiere2024}, who found that the majority (85\;per\;cent) of the QSOFEED sample have $\mathrm{[O\,III]}\lambda5007$ $W_\mathrm{80}$ values that are above what is measured for the stellar populations. Hence, this is a strong indication that the line profiles of our sample are dominated by outflowing-gas emission.

\section{Results}
\label{section: results}

% \subsection{Comparison of electron-density diagnostics}
% \label{section: results: comparison}

\begin{figure}
	\includegraphics[width=\linewidth]{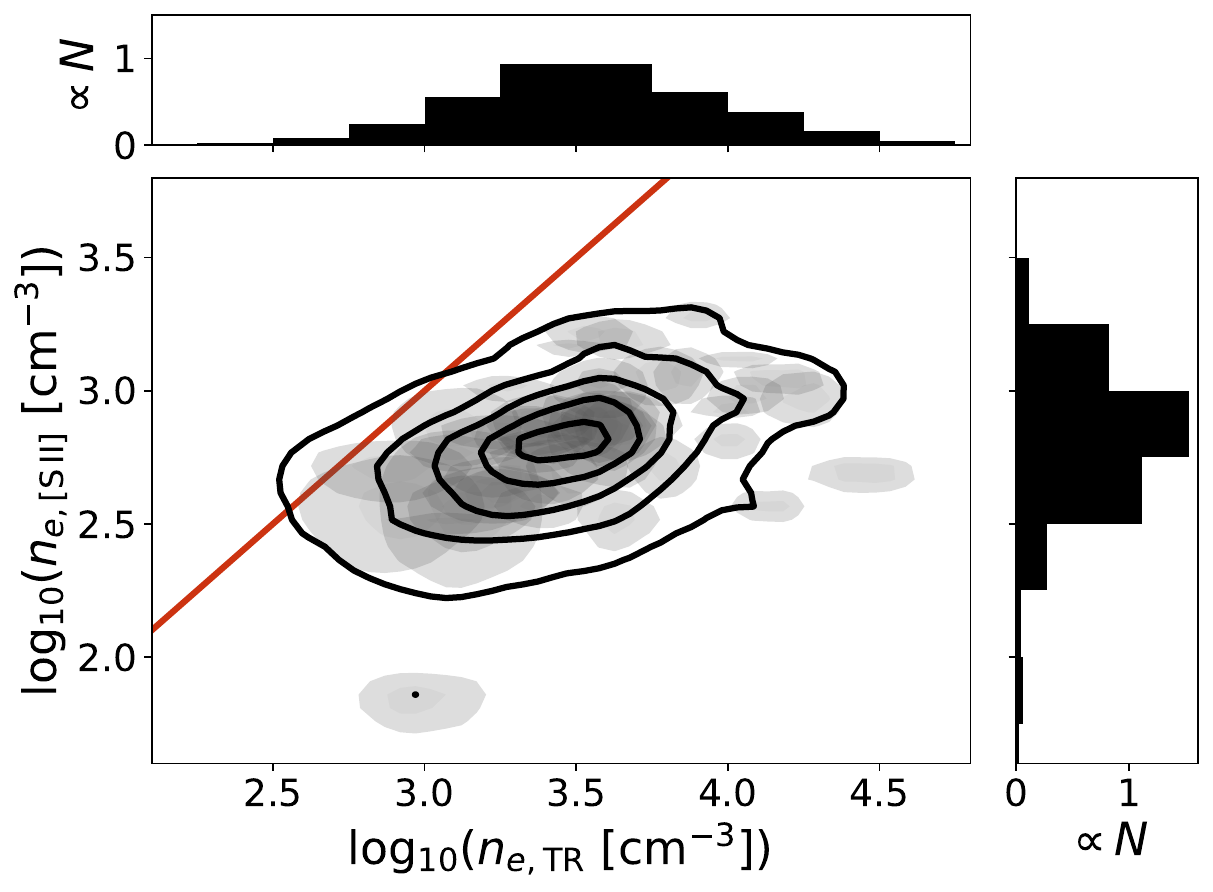}
	\caption{Two-dimensional histogram of electron densities measured by the transauroral-line-technique ($TR$) and [S\,II](6717/6731) flux ratio for the 10,000 Monte Carlo realisations for each object in the QSOFEED sample; one-dimensional histograms for the densities derived from each method are shown above and to the right. The black contours contain 10, 30, 50, 70, and 90\;per\;cent of all realisations for the entire sample, while the grey shaded regions contain 67\;per\;cent of the realisations for individual objects. The solid red line represents the one-to-one relation between densities measured with each technique --- it can be seen that the majority of the realisations fall below this line, indicating that the $TR$ method systematically produces values that are $\sim$0.8\;dex higher than the [S\,II] ratio.}
	\label{fig: 2d_ne_hist}
\end{figure}

In Figure \ref{fig: 2d_ne_hist} we present a two-dimensional histogram of all Monte Carlo realisations for which the randomly-drawn [S\,II](6717/6731) values were within the ratio limits ($0.46<\mathrm{[S\;II](6717/6731)}<1.45$). It can be seen that the distribution of electron-density values is systematically offset towards those measured using the transauroral-line technique, with the peak lying $\sim$0.8\;dex away from the one-to-one line (shown in red) along the $\mathrm{log}_{10}(n_{e, \mathrm{TR}})$ axis. This is a clear indication that the transauroral-line method produces systematically higher densities than the [S\,II](6717/6731) ratio.

\begin{figure}
	\includegraphics[width=\linewidth]{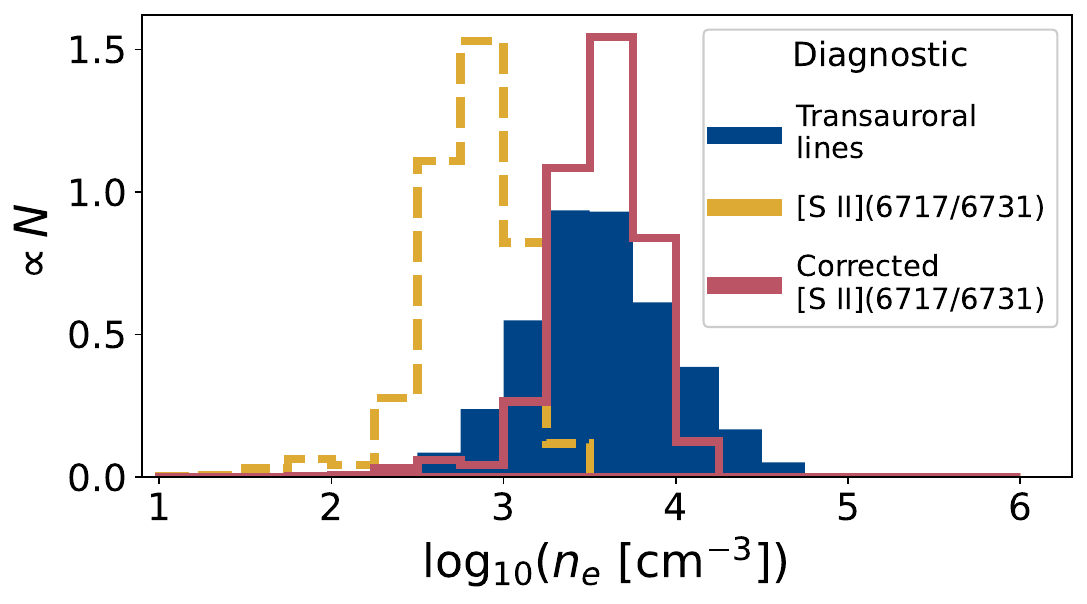}
	\caption{Electron-density histograms produced using different diagnostics with our Monte Carlo approach for the QSOFEED sample: the transauroral-line technique ($TR$; blue solid bars), the [S\,II](6717/6731) ratio without correction (dashed yellow line), and the [S\,II](6717/6731) ratio with a correction factor of 0.75\;dex (solid red line). The former two methods produce distinct electron-density distributions, with values from the $TR$ technique being systematically higher and extending to higher values than can be measured with the [S\,II] ratio. However, values produced using the [S\,II] ratio with our correction factor are statistically consistent with those produced with the transauroral lines for our sample.}
	\label{fig: 1d_ne_hists}
\end{figure}

The one-dimensional electron-density histograms from each technique, shown in Figure \ref{fig: 1d_ne_hists}, clearly show that the values derived from the transauroral-line ratios and the [S\,II](6717/6731) ratio follow different distributions. The distribution for the transauroral-line technique peaks at a higher electron density and extends to higher values ($4.0<\mathrm{log}_{10}(n_e\mathrm{ [cm}^{-3}\mathrm{]})<5.0$) that have no corresponding realisations from the [S\,II](6717/6731) ratio. We quantified this difference by measuring the means of each distribution (and the associated standard error of the mean for each), which we found to be $\mathrm{log}_{10}(n_{e\mathrm{,\;}TR}\mathrm{ [cm}^{-3}\mathrm{]})=3.55\pm0.06$ and $\mathrm{log}_{10}(n_{e\mathrm{,\;[S\,II]}}\mathrm{ [cm}^{-3}\mathrm{]})=2.80\pm0.04$. 

% \subsection{The relationship between the [S\,II] and $TR$-derived density distributions}
% \label{section: results: correction_factor}

To attempt to quantify any potential relationship between the density distributions from the two diagnostics, we added the difference of the means of the distributions ($0.75\pm0.07$\;dex) to the [S\,II]-derived density values. We present the resulting histogram, plotted over the original $TR$-derived densities, in Figure\;\ref{fig: 1d_ne_hists}. 

The resulting `corrected' [S\,II](6717/6731) density distribution appears similar in shape to that derived from the transauroral-line technique. To determine if the distributions differ, we randomly drew 48 samples from each (matching our sample size) 10,000 times and performed Kolmogorov-Smirnov (KS) tests --- we found that 93.1\;per\;cent of the $p$-values from these tests lie above a significance level of 0.05, meaning that we cannot reject the null hypothesis that the samples are drawn from the same parent distribution. Overall, this is evidence that, for our sample, the corrected [S\,II]-derived electron densities are statistically comparable to those derived from the transauroral lines. 

\section{Discussion}
\label{section: discussion}

We have used a Monte Carlo approach to determine electron densities for the warm-ionised gas outflows in the QSOFEED sample of 48 nearby QSO2s using two techniques: the commonly-used [S\,II](6717/6731) ratio and the transauroral-line-ratio ($TR$) method. Our results have shown that the density measurements produced by the [S\,II] ratio are systematically lower than those produced by the $TR$ technique, and that the densities from each follow different distributions. Moreover, we have demonstrated that the difference of the means of these distributions can be used to correct [S\,II]-derived densities, resulting in an electron-density distribution that is consistent with the one produced using the transauroral lines for our sample. In this section, we discuss the physical interpretations of these results, propose a correction factor for densities derived using the [S\,II](6717/6731) ratio, and suggest use cases for the two density diagnostics in future studies of AGN-driven outflows.

\subsection{Physical interpretation of the differences in density produced by the two diagnostics}
\label{section: discussion: interpretation}

As can be clearly seen in the electron-density distributions produced using the [S\,II](6717/6731) ratio and the transauroral-line technique (Figures\;\ref{fig: 2d_ne_hist}\;and\;\ref{fig: 1d_ne_hists}) --- and as was quantified in Section\;\ref{section: results} --- the $TR$-derived densities for our sample are systematically higher by $0.75\pm0.07$\;dex. This can be explained as the transauroral [O\,II]$\lambda\lambda7319,7330$ and [S\,II]$\lambda\lambda4068,4076$ lines tracing denser gas than the [S\,II]$\lambda\lambda6717,6731$ doublet. Such an intepretation is consistent with what would be expected based on the higher critical densities of the transauroral lines (see Appendix\;A in \citealt{HoldenTadhunter2023}) and the locally-optimally-emitting cloud model presented by \citet{Baldwin1995}, in which combinations of clouds of different densities and distances from the central ionising source can explain the observed line strengths in quasar spectra.

In this context, it is important that \citet{Holden2023} used spatially-resolved observations of the outflows in the Seyfert\;2 galaxy IC\;5063 to provide evidence that the transauroral lines are emitted at the same locations as (and with similiar kinematics to) other key diagnostic lines (including [S\,II]$\lambda\lambda6717,6731$ and [O\,III]$\lambda\lambda4959,5007$), indicating that they trace different parts of the same clouds or cloud complexes. This suggests that it is electron density, not distance from the ionising source, that is the dominant parameter for determining the relative strength of commonly-used outflow-diagnostic lines. Therefore, given that the [O\,III]$\lambda\lambda4959,5007$ lines --- which are often used as a kinematic tracer for ionised outflows (e.g. \citealt{Whittle1988, Crenshaw2000_N1068, Das2007, Mullaney2013, Tadhunter2019}) --- have a higher critical density than the $TR$ lines, our results support a scenario in which the transauroral [O\,II]$\lambda\lambda7319,7330$ and [S\,II]$\lambda\lambda4068,4076$ lines trace parts of cloud complexes that have a density in between those that emit the [S\,II]$\lambda\lambda6717,6731$ and [O\,III]$\lambda\lambda4959,5007$ doublets. Thus, we argue that in cases where kinematics have been established with the [O\,III]$\lambda\lambda4959,5007$ doublet, the transauroral lines are a better tracer of the observed outflowing gas than the [S\,II]$\lambda\lambda6717,6731$ doublet. This is supported by the recent findings of \citet{RamosAlmeida2025}, who used mid-infrared spectroscopy of five objects in the QSOFEED sample to show that the densities derived from the flux ratio of the high-critical-density, high-ionisation $\mathrm{[Ne\;V]}14.3, 24.3$\;\textmu m emission lines are comparable to those derived from the transauroral lines.

\subsection{A [S\,II](6717/6731) electron-density correction factor for AGN-driven outflow studies}
\label{section: discussion: correction_factor}

Since the transauroral lines are sensitive to densities that are closer to the critical density of the [O\,III]$\lambda\lambda4959,5007$ doublet, there is a clear motivation to use the $TR$ technique when calculating the properties of AGN-driven outflows. However, a major limitation of the transauroral-line method is that the lines involved are typically faint --- therefore limiting their use beyond deep observations of nearby objects --- and require a wide wavelength coverage (3700--7400\;{\AA} at $z=0$) to detect.

In the case where the transauroral [O\,II]$\lambda\lambda7319,7330$ and [S\,II]$\lambda\lambda4068,4076$ lines are not detected, we propose the use of the difference of the means of the logarithmic [S\,II](6717/6731) and $TR$ electron-density distributions as a correction factor to improve [S\,II] density estimates. In this way, the systematic offset in measured density --- which we interpret as the different lines arising from distinct parts of outflowing clouds --- can be accounted for. As demonstrated in Section\;\ref{section: results}, a correction factor of $0.75\pm0.07$\;dex can be added to [S\,II](6717/6731) electron-density measurements for our sample to produce values that are comparable to those derived from the transauroral-line method. This represents a significant improvement in accuracy over using the [S\,II] ratio alone or simply assuming values of $1.5<\mathrm{log}_{10}(n_e\mathrm{ [cm}^{-3}\mathrm{]})<3.0$, the latter of which is sometimes done (e.g. \citealt{Liu2013, Genzel2014, Fiore2017, Travascio2024, Vayner2024}).

Since we do not separate outflowing gas (which has been observed to have higher densities: \citealt{Holt2011, Rose2018}) from non-outflowing gas, our correction factor is likely an underestimate for AGN-driven outflows due to the saturation of the [S\,II](6717/6731) ratio at high densities (see Figure\;\ref{fig: sii_curve}). To investigate this, we repeated the analysis presented in Section\;\ref{section: electron_densities} using only the broad Gaussian components (full width half maximum > 300\;km\;s$^{-1}$) of the fits which were detected at the $3\sigma$ level in a given Monte Carlo realisation; the line widths of these components are too large to be explained by regular rotation of host galaxies, and hence we interpret them as arising from outflowing gas. The densities derived from the two techniques for these broad components, where possible, are presented in Table\;\ref{tab: ne}; we are able to make [S\,II]-technique measurements for the broad components for 40 objects, and $TR$ measurements for 21 objects. Considering the uncertainties, all of these density values are consistent with those measured from the total line profiles. Moreoever, we found that the difference between the means of the resulting [S\,II] and $TR$ electron-density distributions was $1.01\pm0.13$\;dex --- using this as a correction factor for the [S\,II] measurements once again produced an electron-density distribution that was statistically indistinguishable from that derived from the $TR$ method. This value is consistent within $2\sigma$ to that which we derived using the sum of all components in the emission-line fits (Section\;\ref{section: results}), and the difference between these values ($0.25\pm0.15$\;dex) is far lower than the value of the correction factor itself. We note that blending between individual line components makes it difficult to robustly separate outflowing and non-outflowing emission in our SDSS spectra, which adds additional uncertainty to this approach; future studies with upcoming, higher-spectral-resolution survey spectrographs such as WEAVE \citep{Shoko2023} and 4MOST \citep{deJong2012} will be able to address this. For these reasons, here we favour the use of the value that is derived from the total line profiles ($0.75\pm0.07$\;dex).

For a measured [S\,II] density of $\mathrm{log}_{10}(n_e\mathrm{ [cm}^{-3}\mathrm{]})=3$, applying this correction factor results in an increase in density of approximately a factor of six.  Since derived mass outflow rates and kinetic powers depend inversely on the electron density (see \citealt{Harrison2018}), this corresponds to a factor-of-six decrease in these parameters. This could potentially change the interpretations made regarding the impact that a given ionised outflow has on the star formation of its host galaxy and, more generally, the importance of AGN-driven outflows in galaxy evolution.

\subsection{Defining use cases for the [S\,II](6717/6731) ratio and the transauroral-line technique}
\label{section: discussion: use_cases}

Based on the discussion given in Section\;\ref{section: discussion: interpretation}, we argue that AGN-driven-outflow studies that use the [O\,III]$\lambda\lambda4959,5007$ lines for gas kinematics should use the transauroral-line electron-density diagnostic where possible: in addition to the densities produced by this method being systematically larger than those measured with the [S\,II](6717/6731) doublet, it is also sensitive to a wider range of densities and does not require measurement of individual lines in doublets with small wavelength separations. The latter advantage is particularly relevant for outflow studies, in which the kinematics often lead to complex line profiles and broad line widths, resulting in blending between the $\mathrm{[S\,II]}\lambda6717$ and $\mathrm{[S\,II]}\lambda6731$ lines.

However, in cases where it is not possible to use the transauroral lines, the [S\,II](6717/6731) ratio can be used with the correction factor of $0.75(\pm0.07)$\;dex that we proposed in Section\;\ref{section: discussion: correction_factor} to provide outflow electron-density estimates that are more accurate than using this ratio without correction or assuming a value:
\begin{equation}
	\mathrm{log}_{10}(n_{e\mathrm{,\, outflow}}\mathrm{ [cm}^{-3}\mathrm{]}) = \mathrm{log}_{10}(n_{e\mathrm{,[S\,II]}}\mathrm{ [cm}^{-3}\mathrm{]}) + 0.75(\pm0.07)
\end{equation}
However, care should be taken with this approach since the correction factor is specific to the QSOFEED sample: due to differences in cloud conditions (namely ionisation and density), its value will likely vary for other samples of AGN. Emission-line studies performed with large spectroscopic surveys will be able to derive other sample-specific correction factors and determine the extent of any variation.

Moreover, the [S\,II]-correction-factor approach does not account for the [S\,II](6717/6731) ratio curve becoming asymptotic (Figure\;\ref{fig: sii_curve}), which may still lead to significant underestimations in the case of high-density gas ($\mathrm{log}_{10}(n_e\mathrm{ [cm}^{-3}])>4$). This may be addressed by requiring that any measured [S\,II](6717/6731) ratio values lie far from the limits defined by atomic physics, and taking an upper/lower limit otherwise (as was done by \citealt{Holden2023}, \citealt{Bessiere2024} and \citealt{HoldenTadhunter2025}). Furthermore, use of this ratio should be avoided altogether in cases where the lines show a high degree of blending due to broad and complex line profiles.

In general, an appropriate diagnostic should be chosen based on the gas which is the target of the study. In the case where lower-density, lower-ionisation gas (e.g non-outflowing gas that has not been compressed by an outflow-acceleration mechanism: \citealt{Holt2011, Holden2023}) is the focus, then is likely that the [S\,II](6717/6731) ratio would be appropriate, and perhaps preferable. Conversely, for studies of AGN-driven outflows, the transauroral lines should be used where possible, and the [S\,II](6717/6731) ratio with a correction factor used otherwise. The latter technique will be essential for statistical studies of large numbers of objects for which the transauroral lines are not detected in the majority of cases, such as those enabled by existing and upcoming DESI \citep{Levi2019} and WEAVE-LOFAR \citep{Smith2016} survey observations.

\section{Conclusions}
\label{section: conclusions}

By measuring electron-densities for the warm-ionised gas in 48 nearby AGN that present clear outflow signatures and accounting for various sources of uncertainty, our study has found the following.

\begin{itemize}
	\item Electron densities measured with the transauroral [O\,II]$\lambda\lambda7319,7330$ and [S\,II]$\lambda\lambda4068,4076$ lines are systematically higher than those estimated with the most-commonly-used density diagnostic: the [S\,II](6717/6731) flux ratio. We argue that this offset is due to the transauroral lines being emitted by denser parts of outflowing gas clouds --- which are similiar to those measured by common kinematic tracers --- therefore highlighting their importance for outflow studies. Since outflow masses depend inversely on electron density, this implies that the [S\;II](6717/6731) ratio leads to significant overestimations of the impact of outflows on their host galaxies
	\item To address this systematic offset, we have derived a correction of $\mathrm{log}_{10}(n_{e\mathrm{,\, outflow}}\mathrm{ [cm}^{-3}\mathrm{]}) = \mathrm{log}_{10}(n_{e\mathrm{,[S\,II]}}\mathrm{ [cm}^{-3}\mathrm{]}) + 0.75(\pm0.07)$, the application of which produces electron densities for our sample that are statistically consistent with those measured with the transauroral lines. Thus, for cases in which measurement with the transauroral-line technique is not possible, we propose that this correction can be used to significantly improve the accuracy of [S\,II](6717/6731) electron-density estimates.
\end{itemize}

Overall, our study provides clear use cases for the transauroral-line and [S\,II](6717/6731) techniques and presents a new method to significantly improve the most-commonly-used electron-density diagnostic --- this is crucial for observationally testing if AGN can provide the impact required of them by models of galaxy evolution.

\section*{Acknowledgements}
The authors thank the anonymous referee for their helpful feedback. LRH and DJBS acknowledge support from the UK Science and Technology Facilities Council (STFC) in the form of grant ST/Y001028/1; DJBS and MIA acknowledge support from STFC grant ST/V000624/1; DJBS and SS acknowledge support from STFC grant ST/X508408/1; MIA acknowledges support from the STFC under grant ST/Y000951/1. This work has made use of the University of Hertfordshire's high-performance computing facility (\href{https://uhhpc.herts.ac.uk/}{https://uhhpc.herts.ac.uk/}). CRA and PC acknowledge support from the Agencia Estatal de Investigaci\'on of the Ministerio de Ciencia, Innovaci\'on y Universidades (MCIU/AEI) under the grant ``Tracking active galactic nuclei feedback from parsec to kiloparsec scales'', with reference PID2022$-$141105NB$-$I00 and the European Regional Development Fund (ERDF).

%%%%%%%%%%%%%%%%%%%%%%%%%%%%%%%%%%%%%%%%%%%%%%%%%%
\section*{Data Availability}

The SDSS legacy survey data are available from Version\;2 of the SDSS Data Archive Server (\url{https://das.sdss.org/www/html/das2.html}); supplementary object information is available through the SDSS DR7 Catalog Archive Server (\url{https://cas.sdss.org/dr7/en/}). BOSS data for the remaining objects is available from the SDSS DR10 (\url{https://www.sdss3.org/dr10/data_access/}).



\bibliographystyle{mnras}
\bibliography{precise_outflow_diagnostics} % if your bibtex file is called example.bib

%\begin{thebibliography}{99}
%\end{thebibliography}

%%%%%%%%%%%%%%%%%%%%%%%%%%%%%%%%%%%%%%%%%%%%%%%%%%

%%%%%%%%%%%%%%%%% APPENDICES %%%%%%%%%%%%%%%%%%%%%

\appendix

\section{Stellar continuum subtraction and modelling}
\label{appendix: ppxf}

\begin{figure*}
	\centering
	\includegraphics[width=\linewidth]{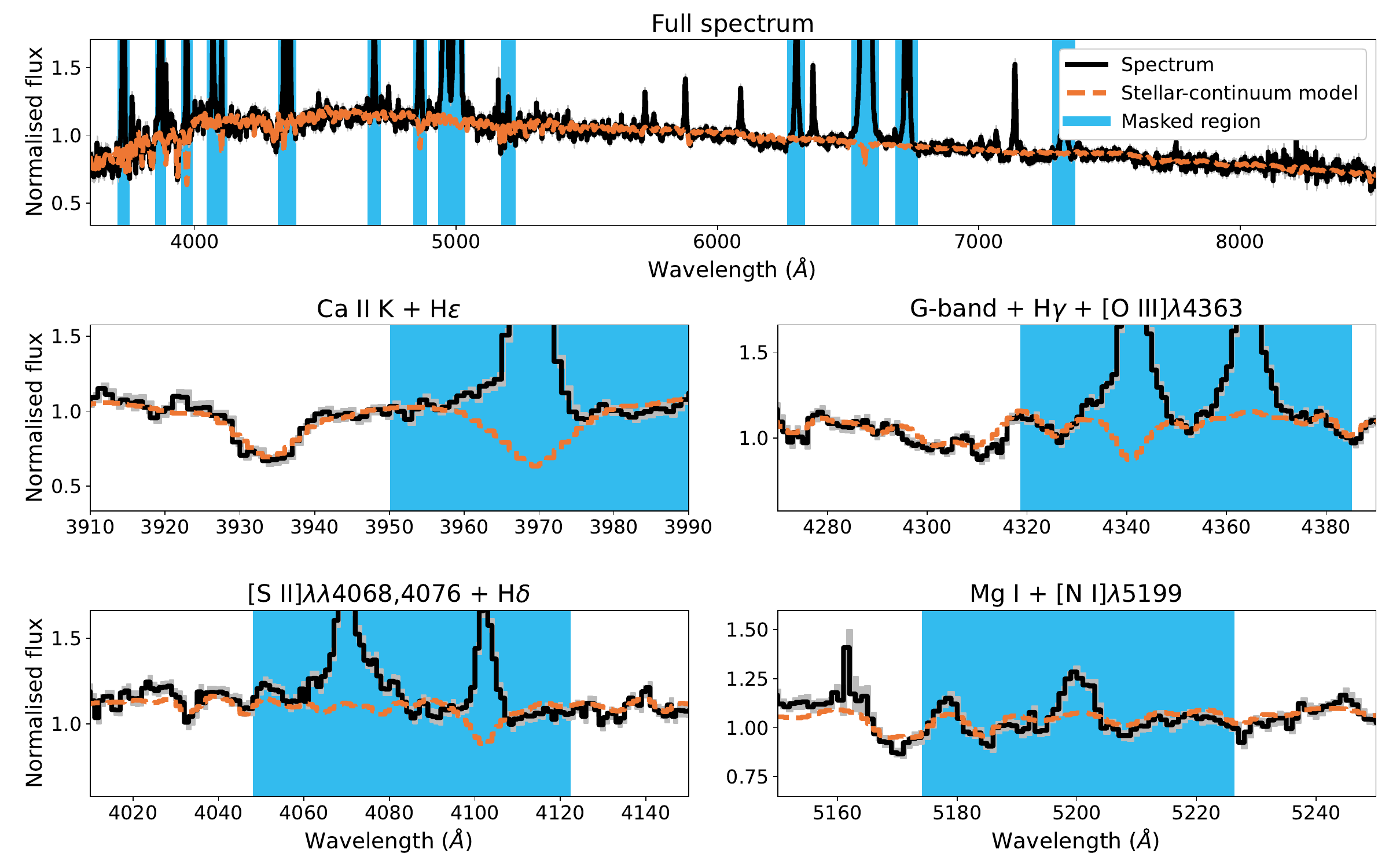}
	\caption{The stellar continuum fit (orange dashed line) to the flux-normalised, de-redshifted spectrum (solid black line; the grey shaded area represents $1\sigma$ flux uncertainties) of the QSO2 J1405+40, produced using \textsc{pPXF} and the \citet{Bruzual2003} stellar templates; regions not included in the fit are shown in shaded blue. The fit across the entire wavelength range in the de-redshifted frame is shown in the top panel, while the other panels show the details of the fit to key stellar absorption features (labelled) and in the region of one of the emission-line doublets used in our analysis ([S\,II]$\lambda\lambda$4068,4076; bottom left panel).}
	\label{fig: stellar_continuum_fit}
\end{figure*}

In order to ensure that the measured fluxes of the faint emission lines involved in our analysis were as accurate as possible, our emission-line fitting routine first modelled the stellar continuum in our spectra using the \textsc{pPXF} code with the \citet{Bruzual2003} stellar templates. The resulting stellar-continuum fits were visually inspected using key stellar spectral features such as the Ca\,II K, Mg\,I, G-band, and Balmer absorption. A representative example of these features in the stellar-continuum fit for one of the objects in our sample is shown in Figure\;\ref{fig: stellar_continuum_fit}, which also demonstrates the typical amount of absorption underlying the transauroral [S\,II]$\lambda\lambda4068,4076$ emission-line doublet.

\section{Line fluxes for the QSOFEED sample}

In Table\;\ref{tab: fluxes}, we present the flux values for the emission lines and emission-line doublets that are used in our analysis, as measured using our spectral fitting routine (Section\;\ref{section: spectral_fitting}). All fluxes are significant at the $3\sigma$ level, with the exception of the [O\,II]($\lambda7319 + \lambda7331$) doublet for J1533+35, which is significant to $1.66\sigma$.

\begin{table*}

        \begin{tabular}{ccccccc}
		Short Name & \parbox{2.7cm}{\centering[O\,II]$(\lambda3726 + \lambda3729)$ \\ Flux \\($\times10^{-16}$\;erg\;s$^{-1}$)} & \parbox{2.7cm}{\centering[S\,II]$(\lambda4068 + \lambda4076)$\\ Flux \\ ($\times10^{-16}$\;erg\;s$^{-1}$)} & \parbox{2.7cm}{\centering[S\,II]$\lambda6717$\\ Flux \\ ($\times10^{-16}$\;erg\;s$^{-1}$)}  & \parbox{2.7cm}{\centering[S\,II]$\lambda6731$\\ Flux \\ ($\times10^{-16}$\;erg\;s$^{-1}$)} & \parbox{2.7cm}{\centering[O\,II]($\lambda7319 + \lambda7331$)\\ Flux \\ ($\times10^{-16}$\;erg\;s$^{-1}$)}\\
        \hline
        J0052-01 & $64.52\pm1.95$ & $4.77\pm0.85$ & $18.35\pm1.03$ & $17.03\pm1.23$ & $4.25\pm1.36$ \\
        J0232-08 & $206.00\pm2.02$ & $16.67\pm0.74$ & $77.25\pm0.97$ & $89.49\pm0.93$ & $14.36\pm2.34$ \\
        J0731+39 & $83.28\pm1.48$ & $15.17\pm0.83$ & $35.29\pm0.45$ & $39.82\pm0.66$ & $16.12\pm1.18$ \\
        J0759+50 & $418.44\pm8.65$ & $91.77\pm4.03$ & $296.43\pm5.18$ & $251.44\pm4.21$ & $149.49\pm7.04$ \\
        J0802+25 & $374.97\pm4.60$ & $30.22\pm1.69$ & $139.96\pm3.05$ & $149.83\pm4.21$ & $34.76\pm3.38$ \\
        J0802+46 & $92.51\pm2.79$ & $12.58\pm0.76$ & $44.64\pm2.66$ & $48.59\pm1.74$ & $16.25\pm2.01$ \\
        J0805+28 & $96.23\pm2.83$ & $11.54\pm1.02$ & $54.18\pm1.49$ & $48.52\pm2.44$ & $16.21\pm2.36$ \\
        J0818+36 & $181.49\pm6.56$ & $11.07\pm1.53$ & $60.99\pm3.43$ & $58.67\pm2.31$ & $19.22\pm3.46$ \\
        J0841+01 & $197.76\pm11.36$ & $5.10\pm1.05$ & $67.16\pm4.17$ & $55.24\pm3.67$ & $6.88\pm2.47$ \\
        J0858+31 & $39.85\pm1.20$ & $2.91\pm0.53$ & $16.03\pm0.79$ & $15.51\pm0.86$ & $5.25\pm0.76$ \\
        J0915+30 & $151.80\pm2.16$ & $8.46\pm0.89$ & $50.39\pm1.33$ & $42.59\pm1.24$ & $8.71\pm1.62$ \\
        J0939+35 & $113.08\pm2.55$ & $2.93\pm0.30$ & $32.03\pm1.04$ & $27.58\pm0.86$ & $9.31\pm0.75$ \\
        J0945+17 & $241.29\pm11.28$ & $16.44\pm1.72$ & $112.66\pm4.38$ & $110.44\pm5.46$ & $32.82\pm3.63$ \\
        J1010+06 & $107.93\pm3.56$ & $39.45\pm2.38$ & $70.10\pm1.37$ & $62.34\pm1.30$ & $64.76\pm4.78$ \\
        J1015+00 & $237.74\pm10.96$ & $8.03\pm1.10$ & $64.43\pm2.67$ & $53.82\pm2.14$ & $9.17\pm2.13$ \\
        J1016+00 & $137.90\pm3.30$ & $5.78\pm0.68$ & $33.78\pm1.06$ & $29.06\pm1.00$ & $4.35\pm1.45$ \\
        J1034+60 & $1069.75\pm34.53$ & $47.30\pm2.53$ & $512.84\pm12.03$ & $474.13\pm12.32$ & $60.15\pm5.38$ \\
        J1036+01 & $134.83\pm1.75$ & $6.41\pm0.39$ & $70.44\pm0.65$ & $58.43\pm0.60$ & $6.70\pm0.70$ \\
        J1100+08 & $277.96\pm5.70$ & $41.60\pm2.40$ & $130.36\pm3.49$ & $123.01\pm3.43$ & $60.21\pm6.28$ \\
        J1137+61 & $204.01\pm2.89$ & $6.33\pm0.19$ & $32.66\pm0.36$ & $30.70\pm0.35$ & $10.36\pm0.45$ \\
        J1152+10 & $367.19\pm9.09$ & $18.20\pm1.69$ & $112.72\pm3.57$ & $109.35\pm3.11$ & $22.07\pm3.97$ \\
        J1157+37 & $94.40\pm3.80$ & $10.70\pm1.09$ & $42.44\pm1.36$ & $50.46\pm1.92$ & $9.12\pm2.29$ \\
        J1200+31 & $355.83\pm21.78$ & $21.70\pm2.31$ & $114.84\pm7.26$ & $113.07\pm6.10$ & $30.07\pm4.63$ \\
        J1218+47 & $96.23\pm7.72$ & $5.69\pm0.91$ & $27.25\pm1.13$ & $24.29\pm1.00$ & $5.87\pm1.10$ \\
        J1223+08 & $72.69\pm1.95$ & $5.77\pm0.73$ & $19.00\pm0.97$ & $20.58\pm1.08$ & $8.06\pm1.70$ \\
        J1238+09 & $102.52\pm6.15$ & $6.66\pm0.86$ & $44.25\pm1.85$ & $39.32\pm1.64$ & $7.15\pm1.54$ \\
        J1241+61 & $111.73\pm4.50$ & $5.31\pm0.93$ & $46.43\pm1.10$ & $44.62\pm1.23$ & $6.97\pm1.39$ \\
        J1244+65 & $109.73\pm7.11$ & $6.97\pm1.10$ & $86.74\pm3.33$ & $82.60\pm3.68$ & $14.96\pm2.75$ \\
        J1300+54 & $302.33\pm4.63$ & $12.42\pm0.91$ & $74.91\pm1.45$ & $70.50\pm1.35$ & $17.96\pm1.90$ \\
        J1316+44 & $139.49\pm5.20$ & $12.18\pm1.35$ & $57.93\pm1.81$ & $48.11\pm1.42$ & $12.90\pm2.37$ \\
        J1347+12 & $132.86\pm6.26$ & $22.30\pm1.14$ & $65.47\pm1.99$ & $67.29\pm3.56$ & $73.82\pm2.63$ \\
        J1356+10 & $345.70\pm2.10$ & $8.57\pm0.52$ & $114.90\pm0.77$ & $83.36\pm0.87$ & $15.91\pm1.11$ \\
        J1356-02 & $42.81\pm2.95$ & $3.49\pm0.83$ & $28.28\pm1.72$ & $27.38\pm1.61$ & $7.36\pm2.23$ \\
        J1405+40 & $170.66\pm2.38$ & $30.77\pm1.40$ & $62.71\pm1.44$ & $67.69\pm1.76$ & $45.05\pm2.29$ \\
        J1430+13 & $897.07\pm20.04$ & $35.53\pm3.98$ & $244.92\pm9.86$ & $242.55\pm9.21$ & $75.32\pm6.70$ \\
        J1436+13 & $226.36\pm10.81$ & $12.98\pm1.21$ & $75.50\pm3.01$ & $90.97\pm3.11$ & $25.47\pm4.06$ \\
        J1437+30 & $320.35\pm6.71$ & $19.29\pm1.72$ & $154.33\pm4.08$ & $162.99\pm3.35$ & $21.91\pm3.40$ \\
        J1440+53 & $2422.26\pm24.64$ & $236.75\pm5.98$ & $687.95\pm8.10$ & $697.64\pm7.81$ & $488.56\pm10.58$ \\
        J1455+32 & $140.88\pm2.22$ & $17.14\pm0.98$ & $46.88\pm1.24$ & $59.18\pm1.23$ & $21.52\pm2.04$ \\
        J1509+04 & $86.39\pm6.71$ & $9.94\pm1.01$ & $61.00\pm2.83$ & $58.35\pm3.71$ & $11.69\pm1.80$ \\
        J1517+33 & $198.36\pm11.16$ & $6.50\pm0.78$ & $108.30\pm7.28$ & $103.85\pm7.09$ & $19.32\pm2.69$ \\
        J1533+35 & $63.11\pm4.71$ & $3.07\pm0.44$ & $24.94\pm2.92$ & $19.80\pm2.13$ & $2.02\pm1.21$ \\
        J1548-01 & $303.05\pm14.56$ & $4.88\pm0.82$ & $58.73\pm2.32$ & $52.43\pm2.12$ & $10.57\pm2.15$ \\
        J1558+35 & $174.70\pm13.12$ & $9.52\pm1.36$ & $60.96\pm2.87$ & $58.96\pm3.22$ & $11.30\pm2.34$ \\
        J1624+33 & $55.96\pm4.36$ & $4.78\pm0.72$ & $29.38\pm1.51$ & $30.96\pm1.65$ & $6.90\pm1.87$ \\
        J1653+23 & $358.55\pm9.81$ & $12.99\pm1.50$ & $99.55\pm2.13$ & $88.03\pm2.42$ & $18.68\pm2.44$ \\
        J1713+57 & $187.28\pm4.36$ & $33.79\pm2.64$ & $90.28\pm2.99$ & $93.79\pm2.91$ & $45.29\pm5.42$ \\
        J2154+11 & $142.59\pm2.55$ & $10.87\pm0.85$ & $68.20\pm1.52$ & $62.45\pm1.22$ & $8.63\pm1.46$ \\
        \end{tabular}
	\caption{Total fluxes of doublets used in the $TR$ ratios (Equations\;\ref{eq: tr_oii} \& \ref{eq: tr_sii}) and the fluxes of the [S\,II]$\lambda6717$ and [S\,II]$\lambda6731$ lines (used to calculate the [S\,II](6717/6731) ratio: Section\;\ref{section: electron_densities: sii}).}
	\label{tab: fluxes}
\end{table*}

\section{An analytic expression for transauroral-line-derived electron densities}
\label{appendix: analytic_tr_ne}

Our Monte Carlo approach to electron-density measurement required a computationally-efficient way of determining electron densities from the grid of transauroral-line-ratio ($TR$) values (Figure\;\ref{fig: tr_grid}) for different photoionisation-model parameters. Therefore, here we derive an analytic expression for deriving electron densities in terms of $TR(\mathrm{[O\,II]})$ and $TR(\mathrm{[S\,II]})$ (Equations\;\ref{eq: tr_oii} and \ref{eq: tr_sii}).

\begin{figure}
	\centering
	\includegraphics[width=1\linewidth]{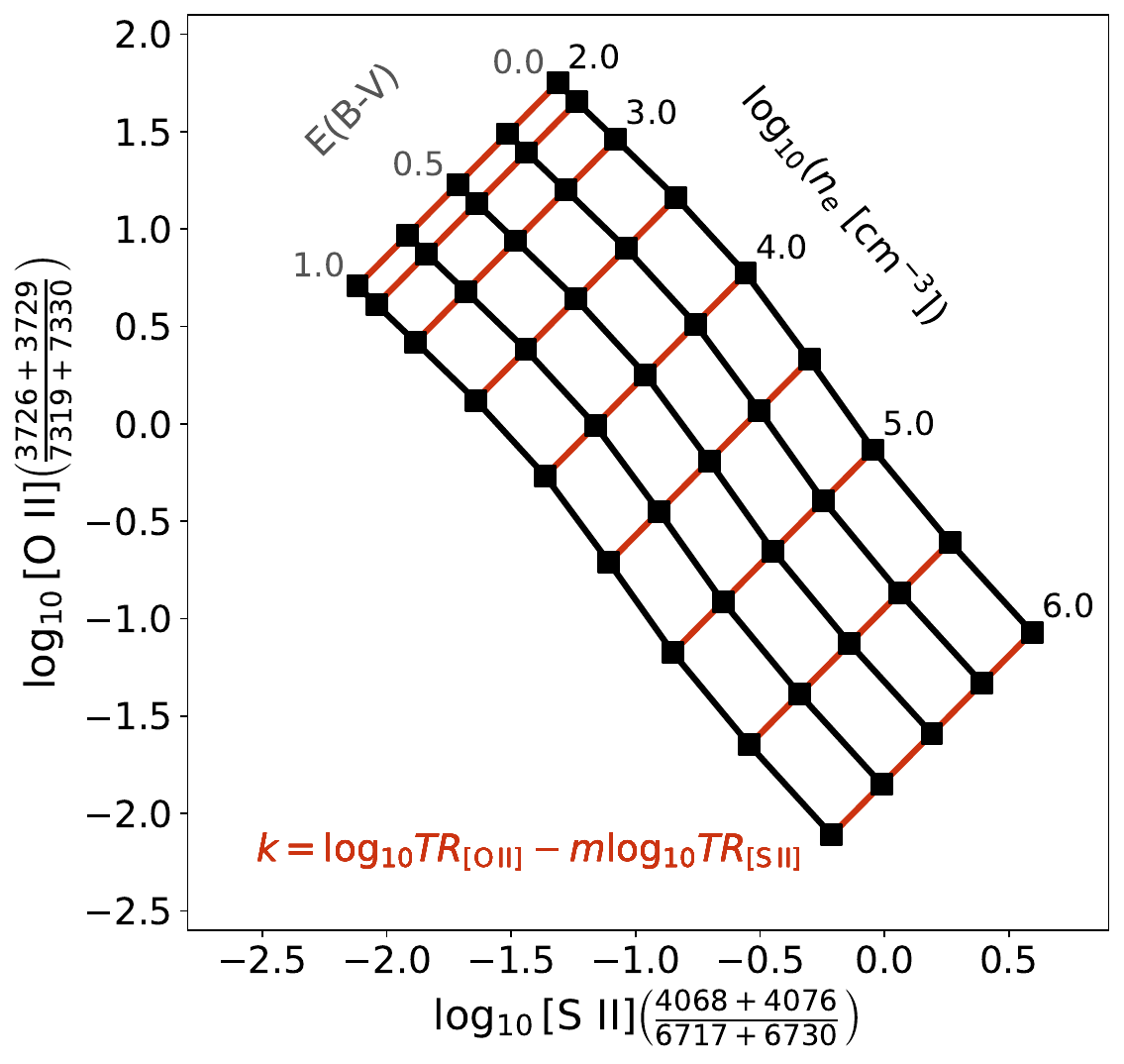}
	\caption{Transauroral-line-ratio ($TR$) grid consisting of the line ratios predicted from photoionisation models for gas of different electron densities ($2.0<\mathrm{log}_{10}(n_e\mathrm{[cm}^{-3}])<6.0$; labelled) and extinction values ($0.0<E(B-V)<1.0$; labelled), for a spectral index of $\alpha=1.5$, an ionisation parameter of $\mathrm{log}U=-2.50$, and solar-metallicity gas. Lines of constant electron density are highlighted in red; a polynomial expression for these lines is given in the bottom left.}
	\label{fig: tr_grid_appendix}
\end{figure}

\begin{figure}
	\centering
	\includegraphics[width=1\linewidth]{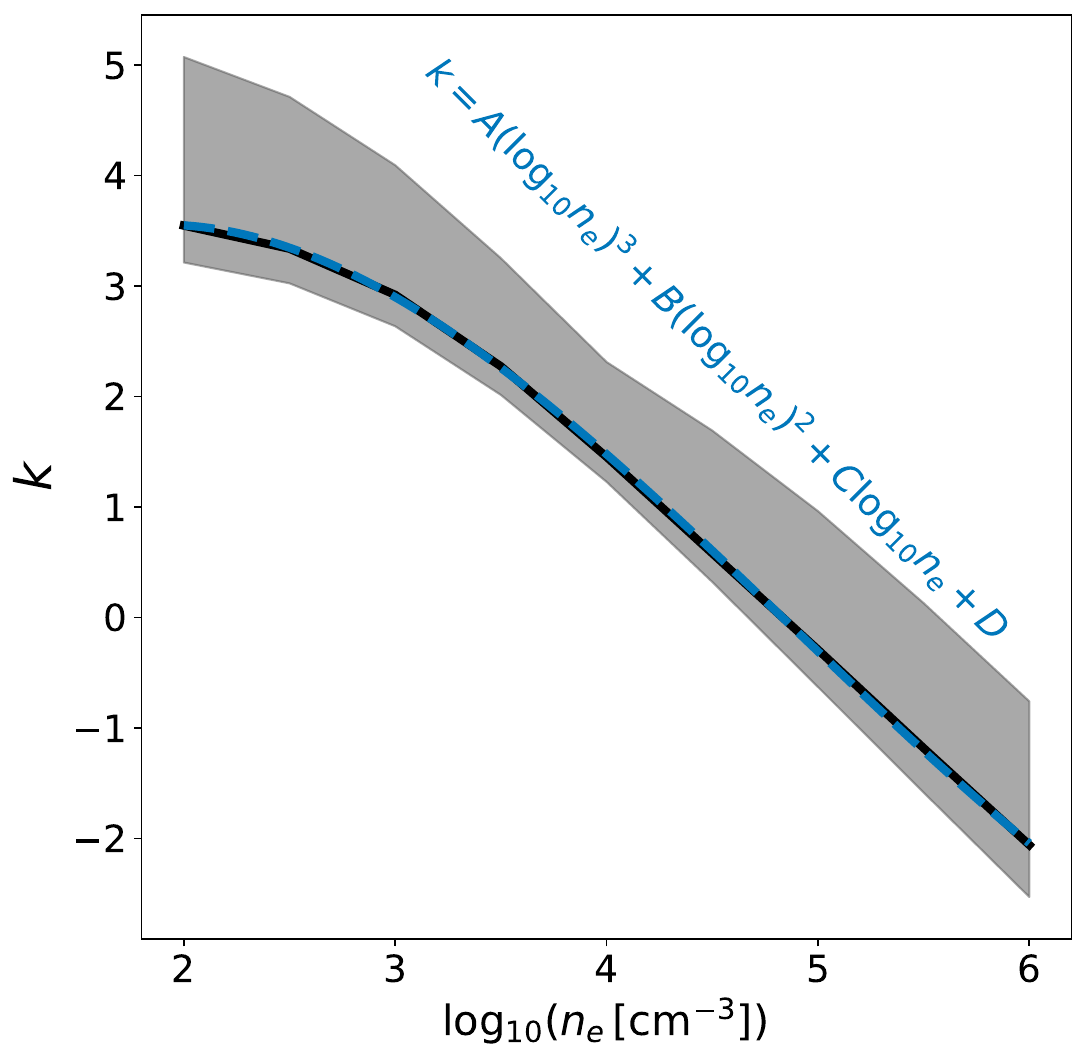}
	\caption{Variation of the constant $k$ (the $TR(\mathrm{[O\,II]})$-axis intercept in Figure\;\ref{fig: tr_grid_appendix}) with electron density, as determined by fitting first-order polynomials to the constant-density lines of the TR grid. The shaded grey region shows the range of values across all photoionisation model parameters that we consider in this work, while the black line is for an example spectral index of $\alpha=1.5$, ionisation parameter $\mathrm{log}U=-3.00$, and solar metallicity gas; the blue dashed line shows a third-order polynomial fit to this line, the general expression for which is given in the top right.}
	\label{fig: ne_c_variation}
\end{figure}

We begin by noting that the lines of constant electron density on a given $TR$ grid (highlighted in red for an example grid in Figure\;\ref{fig: tr_grid_appendix}) are straight, and can be described by a first-order polynomial of the form
\begin{equation}
	\mathrm{log}_{10}TR(\mathrm{[O\,II]}) = m\mathrm{log}_{10}TR(\mathrm{[S\,II]} + k
	\label{eq: tr_c_m}
\end{equation}
where $m$ and $k$ are the gradient and $TR(\mathrm{[O\,II]})$-axis intercept, respectively. While the value of $m=1.293$ for any grid is constant, the value of $k$ depends on the electron density. To evaluate this relationship, we fit the constant-density lines with a first-order polynomial and show the resulting variation of $k$ with $\mathrm{log}_{10}n_e$ in Figure\;\ref{fig: ne_c_variation}. This relationship can be seen to be well described by a third-order polynomial,
\begin{equation}
	k = A\mathrm{log}_{10}n_e^3 + B\mathrm{log}_{10}n_e^2 + C\mathrm{log}_{10}n_e + D,
	\label{eq: c_ne}
\end{equation}
where $A$, $B$, $C$, and $D$ are constants. Combining Equations\;\ref{eq: tr_c_m} and \ref{eq: c_ne} results in an expression for $\mathrm{log}_{10}n_e$ in terms of $TR(\mathrm{[O\,II]})$ and $TR(\mathrm{[S\,II]})$, which we present in Section\;\ref{section: electron_densities: tr} as Equation\;\ref{eq: tr_ne}.

The values of the constants used in this expression depend on the shape and position of the $TR$ grid, which itself depends on the parameters of the photoionisation model used to generate it; no straightforward relationship exists between the constants and model parameters. Therefore, to account for this when calculating electron densities with our Monte Carlo method, we generated a range of photoionisation models with the \textsc{Cloudy} photoionisation code with parameters covering a reasonable range of values expected for the ISM. We varied the 0.025\;nm -- 10\;\textmu m spectral index in the range of $1.0<\alpha<2.0$, which is informed by the results of photoionisation modelling for the NLRs of nearby AGN (e.g. \citealt{Ferland1983, Robinson1987}), ionisation parameters in the range $-4.00<\mathrm{log}U<-2.00$ (covering the typical range of values used in photoionisation modelling of AGN-driven outflows e.g. \citealt{Baron2019b, Revalski2021}), and metallicities in the range $0.5<Z<2.0$\;$Z_\odot$. We present the values of the constants $A$, $B$, $C$, and $D$ for different combinations of photoionisation model parameters in Table\;\ref{tab: tr_expression_constants}.

\begin{table*}
	\begin{tabular}{ccccccc}
	$\alpha$ & log$U$ & $Z$ & $A$ & $B$ & $C$ & $D$ \\
	\hline
	1.0 & -2.00 & 0.5 & 0.017 & -0.441 & 1.594 & 1.913 \\
	1.0 & -2.00 & 1.0 & 0.021 & -0.499 & 1.750 & 1.847 \\
	1.0 & -2.00 & 1.5 & 0.025 & -0.540 & 1.847 & 1.862 \\
	1.0 & -2.00 & 2.0 & 0.026 & -0.552 & 1.838 & 2.007 \\
	1.0 & -2.50 & 0.5 & 0.021 & -0.478 & 1.612 & 1.904 \\
	1.0 & -2.50 & 1.0 & 0.025 & -0.521 & 1.699 & 1.953 \\
	1.0 & -2.50 & 1.5 & 0.028 & -0.555 & 1.762 & 2.045 \\
	1.0 & -2.50 & 2.0 & 0.030 & -0.578 & 1.783 & 2.195 \\
	1.0 & -3.00 & 0.5 & 0.028 & -0.533 & 1.663 & 1.914 \\
	1.0 & -3.00 & 1.0 & 0.034 & -0.604 & 1.817 & 1.968 \\
	1.0 & -3.00 & 1.5 & 0.039 & -0.660 & 1.940 & 2.066 \\
	1.0 & -3.00 & 2.0 & 0.043 & -0.702 & 2.011 & 2.226 \\
	1.0 & -3.50 & 0.5 & 0.045 & -0.719 & 2.214 & 1.469 \\
	1.0 & -3.50 & 1.0 & 0.059 & -0.882 & 2.710 & 1.209 \\
	1.0 & -3.50 & 1.5 & 0.065 & -0.965 & 2.934 & 1.252 \\
	1.0 & -3.50 & 2.0 & 0.069 & -1.009 & 3.017 & 1.455 \\
	1.0 & -4.00 & 0.5 & 0.058 & -0.884 & 2.824 & 0.930 \\
	1.0 & -4.00 & 1.0 & 0.070 & -1.033 & 3.296 & 0.741 \\
	1.0 & -4.00 & 1.5 & 0.074 & -1.088 & 3.419 & 0.976 \\
	1.0 & -4.00 & 2.0 & 0.074 & -1.095 & 3.352 & 1.447 \\
	    &       &     &       &        &       &       \\
	1.5 & -2.00 & 0.5 & 0.026 & -0.518 & 1.606 & 1.956 \\
	1.5 & -2.00 & 1.0 & 0.031 & -0.568 & 1.676 & 2.100 \\
	1.5 & -2.00 & 1.5 & 0.036 & -0.623 & 1.773 & 2.234 \\
	1.5 & -2.00 & 2.0 & 0.041 & -0.670 & 1.847 & 2.402 \\
	1.5 & -2.50 & 0.5 & 0.038 & -0.641 & 1.926 & 1.679 \\
	1.5 & -2.50 & 1.0 & 0.049 & -0.763 & 2.251 & 1.622 \\
	1.5 & -2.50 & 1.5 & 0.056 & -0.842 & 2.445 & 1.698 \\
	1.5 & -2.50 & 2.0 & 0.061 & -0.906 & 2.583 & 1.836 \\
	1.5 & -3.00 & 0.5 & 0.058 & -0.873 & 2.678 & 0.973 \\
	1.5 & -3.00 & 1.0 & 0.068 & -1.001 & 3.056 & 0.893 \\
	1.5 & -3.00 & 1.5 & 0.074 & -1.064 & 3.202 & 1.060 \\
	1.5 & -3.00 & 2.0 & 0.076 & -1.088 & 3.201 & 1.391 \\
	1.5 & -3.50 & 0.5 & 0.069 & -1.014 & 3.186 & 0.530 \\
	1.5 & -3.50 & 1.0 & 0.076 & -1.115 & 3.485 & 0.568 \\
	1.5 & -3.50 & 1.5 & 0.079 & -1.146 & 3.519 & 0.906 \\
	1.5 & -3.50 & 2.0 & 0.079 & -1.149 & 3.432 & 1.393 \\
	1.5 & -4.00 & 0.5 & 0.069 & -1.032 & 3.308 & 0.507 \\
	1.5 & -4.00 & 1.0 & 0.075 & -1.114 & 3.544 & 0.678 \\
	1.5 & -4.00 & 1.5 & 0.075 & -1.116 & 3.448 & 1.263 \\
	1.5 & -4.00 & 2.0 & 0.074 & -1.094 & 3.235 & 2.002 \\
	    &       &     &       &        &       &       \\
	2.0 & -2.00 & 0.5 & 0.064 & -0.923 & 2.780 & 0.848 \\
	2.0 & -2.00 & 1.0 & 0.084 & -1.170 & 3.592 & 0.275 \\
	2.0 & -2.00 & 1.5 & 0.094 & -1.304 & 4.030 & 0.082 \\
	2.0 & -2.00 & 2.0 & 0.098 & -1.360 & 4.198 & 0.168 \\
	2.0 & -2.50 & 0.5 & 0.075 & -1.067 & 3.296 & 0.371 \\
	2.0 & -2.50 & 1.0 & 0.085 & -1.204 & 3.731 & 0.252 \\
	2.0 & -2.50 & 1.5 & 0.089 & -1.259 & 3.869 & 0.444 \\
	2.0 & -2.50 & 2.0 & 0.092 & -1.296 & 3.936 & 0.701 \\
	2.0 & -3.00 & 0.5 & 0.078 & -1.126 & 3.532 & 0.191 \\
	2.0 & -3.00 & 1.0 & 0.082 & -1.191 & 3.699 & 0.423 \\
	2.0 & -3.00 & 1.5 & 0.084 & -1.213 & 3.684 & 0.854 \\
	2.0 & -3.00 & 2.0 & 0.085 & -1.212 & 3.575 & 1.388 \\
	2.0 & -3.50 & 0.5 & 0.079 & -1.155 & 3.662 & 0.143 \\
	2.0 & -3.50 & 1.0 & 0.082 & -1.200 & 3.750 & 0.515 \\
	2.0 & -3.50 & 1.5 & 0.082 & -1.193 & 3.610 & 1.156 \\
	2.0 & -3.50 & 2.0 & 0.082 & -1.181 & 3.435 & 1.839 \\
	2.0 & -4.00 & 0.5 & 0.075 & -1.119 & 3.590 & 0.334 \\
	2.0 & -4.00 & 1.0 & 0.077 & -1.150 & 3.605 & 0.878 \\
	2.0 & -4.00 & 1.5 & 0.076 & -1.124 & 3.357 & 1.761 \\
	2.0 & -4.00 & 2.0 & 0.073 & -1.069 & 2.967 & 2.832 \\
	\end{tabular}
	\caption{Values for the constants $A$, $B$, $C$, $D$, and $m$ --- used with Equation\;\ref{eq: tr_ne} to calculate electron densities --- for photoionisation models with various combinations of 0.025\;nm -- 10\;{\textmu}m spectral index ($\alpha$), ionisation parameter ($\mathrm{log}U$), and metallicity ($Z$). The value of the constant $m$ is the same for all grids: $m=1.293$.}
	\label{tab: tr_expression_constants}
\end{table*}

\section{Electron density values for the QSOFEED sample}
\label{appendix: ne_table}

In Table\;\ref{tab: ne}, we present the electron density values for each object in our sample, derived using our Monte Carlo approach with the transauroral-line-ratio ($TR$) and [S\,II](6717/6731) ratio techniques (see Section\;\ref{section: electron_densities}). The values were taken to be the 50th percentile of the electron-density distribution for each object, while the quoted upper and lower uncertainties are the 16th and 84th percentiles. Individual $TR$ electron densities for all objects in the sample are consistent within $3\sigma$ to those measured using the transauroral-line technique (where possible) for the QSOFEED sample by \citet{Bessiere2024}, with the majority (80.0\;per\;cent) being consistent within $1\sigma$. Note that \citet{Bessiere2024} presented transauroral-line-derived electron densities only in cases where the lines were clearly detected, and values from the [S\,II](6717/6731) ratio where this was not possible.
\begin{table*}
	\renewcommand*{\arraystretch}{1.31}
	\centering
        \begin{tabular}{cccccc}
		Short Name & SDSS ID & $\log_{10}(n_{e,\mathrm{TR}}\;[\mathrm{cm}^{-3}])$ & $\log_{10}(n_{e,\mathrm{[S\,II]}}\;[\mathrm{cm}^{-3}])$ & $\log_{10}(n_{e,\mathrm{TR\,{outflow}}}\;[\mathrm{cm}^{-3}])$ & $\log_{10}(n_{e,\mathrm{[S\,II],\,{outflow}}}\;[\mathrm{cm}^{-3}])$ \\
        \hline
       J0052-01 & J005230.59-011548.4 & 3.60$^{+0.21}_{-0.19}$ & 2.77$^{+0.17}_{-0.20}$ & 3.63$^{+0.22}_{-0.21}$ & 2.86$^{+0.18}_{-0.23}$ \\
       J0232-08 & J023224.24-081140.2 & 3.50$^{+0.22}_{-0.16}$ & 3.15$^{+0.04}_{-0.03}$ & 3.51$^{+0.23}_{-0.15}$ & 3.14$^{+0.03}_{-0.03}$ \\
       J0731+39 & J073142.37+392623.7 & 4.06$^{+0.18}_{-0.15}$ & 3.11$^{+0.04}_{-0.04}$ & 4.62$^{+0.19}_{-0.18}$ & 4.22$^{+0.22}_{-0.21}$ \\
       J0759+50 & J075940.95+505023.9 & 4.15$^{+0.19}_{-0.14}$ & 2.57$^{+0.06}_{-0.07}$ & 4.31$^{+0.18}_{-0.16}$ & 2.32$^{+0.11}_{-0.13}$ \\
       J0802+25 & J080252.92+255255.5 & 3.61$^{+0.20}_{-0.15}$ & 3.03$^{+0.06}_{-0.06}$ & 4.15$^{+0.18}_{-0.16}$ & 4.02$^{+0.26}_{-0.23}$ \\
       J0802+46 & J080224.34+464300.7 & 3.89$^{+0.19}_{-0.15}$ & 3.06$^{+0.10}_{-0.12}$ & 4.14$^{+0.19}_{-0.16}$ & 3.42$^{+0.33}_{-0.28}$ \\
       J0805+28 & J080523.29+281815.7 & 3.81$^{+0.20}_{-0.15}$ & 2.70$^{+0.11}_{-0.14}$ & 3.91$^{+0.19}_{-0.16}$ & 2.53$^{+0.22}_{-0.34}$ \\
       J0818+36 & J081842.35+360409.6 & 3.61$^{+0.19}_{-0.18}$ & 2.85$^{+0.12}_{-0.14}$ & ---- & 3.35$^{+0.34}_{-0.30}$ \\
       J0841+01 & J084135.09+010156.3 & 2.86$^{+0.31}_{-0.34}$ & 2.49$^{+0.22}_{-0.33}$ & ---- & ---- \\
       J0858+31 & J085810.63+312136.2 & 3.67$^{+0.20}_{-0.17}$ & 2.86$^{+0.12}_{-0.15}$ & ---- & 2.95$^{+0.30}_{-0.41}$ \\
       J0915+30 & J091544.18+300922.0 & 3.41$^{+0.21}_{-0.19}$ & 2.56$^{+0.09}_{-0.12}$ & ---- & 2.18$^{+0.34}_{-0.52}$ \\
       J0939+35 & J093952.75+355358.9 & 3.27$^{+0.22}_{-0.20}$ & 2.61$^{+0.10}_{-0.13}$ & ---- & 2.55$^{+0.39}_{-0.54}$ \\
       J0945+17 & J094521.33+173753.2 & 3.59$^{+0.21}_{-0.16}$ & 2.88$^{+0.11}_{-0.12}$ & 3.88$^{+0.20}_{-0.16}$ & 3.03$^{+0.21}_{-0.25}$ \\
       J1010+06 & J101043.36+061201.4 & 4.46$^{+0.17}_{-0.17}$ & 2.68$^{+0.06}_{-0.07}$ & 4.79$^{+0.22}_{-0.18}$ & 1.78$^{+0.32}_{-0.51}$ \\
       J1015+00 & J101536.21+005459.4 & 3.14$^{+0.24}_{-0.24}$ & 2.53$^{+0.14}_{-0.19}$ & ---- & 2.86$^{+0.24}_{-0.30}$ \\
       J1016+00 & J101653.82+002857.2 & 3.21$^{+0.23}_{-0.25}$ & 2.60$^{+0.11}_{-0.13}$ & ---- & ---- \\
       J1034+60 & J103408.59+600152.2 & 3.11$^{+0.23}_{-0.22}$ & 2.76$^{+0.08}_{-0.07}$ & ---- & 1.54$^{+0.32}_{-0.54}$ \\
       J1036+01 & J103600.37+013653.5 & 3.08$^{+0.24}_{-0.24}$ & 2.50$^{+0.04}_{-0.04}$ & ---- & ---- \\
       J1100+08 & J110012.39+084616.3 & 4.02$^{+0.18}_{-0.15}$ & 2.81$^{+0.07}_{-0.08}$ & 4.17$^{+0.19}_{-0.15}$ & 2.51$^{+0.17}_{-0.22}$ \\
       J1137+61 & J113721.36+612001.1 & 3.41$^{+0.22}_{-0.16}$ & 2.80$^{+0.04}_{-0.04}$ & ---- & ---- \\
       J1152+10 & J115245.66+101623.8 & 3.38$^{+0.21}_{-0.18}$ & 2.86$^{+0.08}_{-0.08}$ & ---- & 3.11$^{+0.28}_{-0.29}$ \\
       J1157+37 & J115759.50+370738.2 & 3.66$^{+0.20}_{-0.17}$ & 3.19$^{+0.08}_{-0.07}$ & ---- & 3.54$^{+0.17}_{-0.15}$ \\
       J1200+31 & J120041.39+314746.2 & 3.55$^{+0.20}_{-0.17}$ & 2.89$^{+0.13}_{-0.16}$ & ---- & ---- \\
       J1218+47 & J121839.40+470627.7 & 3.51$^{+0.20}_{-0.18}$ & 2.69$^{+0.12}_{-0.15}$ & ---- & 3.48$^{+0.49}_{-0.61}$ \\
       J1223+08 & J122341.47+080651.3 & 3.79$^{+0.19}_{-0.17}$ & 3.05$^{+0.11}_{-0.12}$ & ---- & 3.47$^{+0.33}_{-0.49}$ \\
       J1238+09 & J123843.44+092736.6 & 3.42$^{+0.21}_{-0.19}$ & 2.68$^{+0.12}_{-0.15}$ & ---- & ---- \\
       J1241+61 & J124136.22+614043.4 & 3.25$^{+0.22}_{-0.23}$ & 2.84$^{+0.07}_{-0.07}$ & ---- & 2.97$^{+0.17}_{-0.20}$ \\
       J1244+65 & J124406.61+652925.2 & 3.35$^{+0.23}_{-0.20}$ & 2.82$^{+0.11}_{-0.12}$ & 3.41$^{+0.24}_{-0.26}$ & 3.02$^{+0.21}_{-0.25}$ \\
       J1300+54 & J130038.09+545436.8 & 3.39$^{+0.22}_{-0.17}$ & 2.80$^{+0.06}_{-0.06}$ & ---- & ---- \\
       J1316+44 & J131639.74+445235.0 & 3.65$^{+0.20}_{-0.16}$ & 2.51$^{+0.11}_{-0.14}$ & ---- & 2.67$^{+0.43}_{-0.55}$ \\
       J1347+12 & J134733.36+121724.3 & 4.27$^{+0.17}_{-0.16}$ & 2.96$^{+0.10}_{-0.11}$ & 4.56$^{+0.18}_{-0.17}$ & 3.35$^{+0.15}_{-0.16}$ \\
       J1356+10 & J135646.10+102609.0 & 2.97$^{+0.26}_{-0.27}$ & 1.81$^{+0.12}_{-0.15}$ & 2.94$^{+0.28}_{-0.29}$ & 1.21$^{+0.29}_{-0.38}$ \\
       J1356-02 & J135617.79-023101.5 & 3.59$^{+0.22}_{-0.20}$ & 2.85$^{+0.15}_{-0.16}$ & ---- & 2.53$^{+0.39}_{-0.53}$ \\
       J1405+40 & J140541.21+402632.6 & 4.19$^{+0.18}_{-0.15}$ & 3.04$^{+0.06}_{-0.06}$ & 4.59$^{+0.19}_{-0.17}$ & 2.93$^{+0.22}_{-0.29}$ \\
       J1430+13 & J143029.88+133912.0 & 3.44$^{+0.21}_{-0.18}$ & 2.90$^{+0.09}_{-0.11}$ & 3.13$^{+0.25}_{-0.26}$ & 2.65$^{+0.13}_{-0.15}$ \\
       J1436+13 & J143607.21+492858.6 & 3.56$^{+0.20}_{-0.17}$ & 3.21$^{+0.08}_{-0.08}$ & 3.45$^{+0.21}_{-0.18}$ & 3.62$^{+0.11}_{-0.11}$ \\
       J1437+30 & J143737.85+301101.1 & 3.30$^{+0.22}_{-0.20}$ & 3.01$^{+0.05}_{-0.06}$ & 4.02$^{+0.20}_{-0.16}$ & 3.87$^{+0.32}_{-0.25}$ \\
       J1440+53 & J144038.10+533015.9 & 4.01$^{+0.19}_{-0.14}$ & 2.94$^{+0.03}_{-0.04}$ & 4.32$^{+0.18}_{-0.16}$ & 2.56$^{+0.09}_{-0.11}$ \\
       J1455+32 & J145519.41+322601.8 & 3.91$^{+0.19}_{-0.14}$ & 3.28$^{+0.06}_{-0.05}$ & 4.12$^{+0.19}_{-0.15}$ & 3.83$^{+0.14}_{-0.12}$ \\
       J1509+04 & J150904.22+043441.8 & 3.64$^{+0.20}_{-0.16}$ & 2.83$^{+0.14}_{-0.16}$ & 4.15$^{+0.19}_{-0.17}$ & 2.77$^{+0.32}_{-0.49}$ \\
       J1517+33 & J151709.20+335324.7 & 3.10$^{+0.25}_{-0.25}$ & 2.84$^{+0.16}_{-0.20}$ & ---- & 3.11$^{+0.21}_{-0.22}$ \\
       J1533+35 & J153338.03+355708.1 & 3.09$^{+0.27}_{-0.33}$ & 2.49$^{+0.33}_{-0.49}$ & ---- & ---- \\
       J1548-01 & J154832.37-010811.8 & 2.88$^{+0.29}_{-0.33}$ & 2.69$^{+0.12}_{-0.14}$ & ---- & 2.17$^{+0.31}_{-0.49}$ \\
       J1558+35 & J155829.36+351328.6 & 3.39$^{+0.22}_{-0.20}$ & 2.86$^{+0.12}_{-0.15}$ & ---- & 3.11$^{+0.19}_{-0.22}$ \\
       J1624+33 & J162436.40+334406.7 & 3.59$^{+0.21}_{-0.19}$ & 3.00$^{+0.12}_{-0.12}$ & ---- & 2.65$^{+0.32}_{-0.46}$ \\
       J1653+23 & J165315.05+234942.9 & 3.26$^{+0.22}_{-0.21}$ & 2.67$^{+0.07}_{-0.09}$ & ---- & 2.65$^{+0.26}_{-0.40}$ \\
       J1713+57 & J171350.32+572954.9 & 4.08$^{+0.18}_{-0.15}$ & 2.98$^{+0.08}_{-0.08}$ & 4.19$^{+0.19}_{-0.15}$ & 3.13$^{+0.09}_{-0.10}$ \\
       J2154+11 & J215425.74+113129.4 & 3.39$^{+0.21}_{-0.18}$ & 2.75$^{+0.06}_{-0.07}$ & ---- & 1.93$^{+0.33}_{-0.55}$ \\
        \end{tabular}
	\caption{Electron-density values for the QSOFEED sample, produced using our Monte Carlo approach with the transauroral-line technique ($TR$; see Section\;\ref{section: electron_densities: tr}) and [S\,II](6717/6731) ratio (Section\;\ref{section: electron_densities: sii}.). The leftmost electron-density columns give the densities derived using the total emission-line profiles (which we use in our analysis: see Section\;\ref{section: discussion: correction_factor}), while the rightmost electron-density columns (labelled with the `outflow' subscript) are those derived using the broad Gaussian components ($\mathrm{FWHM}>300$\;km\;s$^{-1}$) only.}
	\label{tab: ne}

\end{table*}
%%%%%%%%%%%%%%%%%%%%%%%%%%%%%%%%%%%%%%%%%%%%%%%%%%

% Don't change these lines
\bsp	% typesetting comment
\label{lastpage}
\end{document}